\documentclass[journal,draftclsnofoot,onecolumn,12pt]{IEEEtran}
\usepackage[T1]{fontenc}
\IEEEoverridecommandlockouts
\usepackage{cite}
\usepackage{amsmath,amssymb,amsfonts}
\usepackage{bm}
\usepackage{algorithmic}
\usepackage{graphicx}
\usepackage{subfigure}
\usepackage{textcomp}
\usepackage{xcolor}
\usepackage{threeparttable}
\usepackage{multirow}
\usepackage{verbatim}
\usepackage{url}
\usepackage{hyperref}
\usepackage{comment}
\def\BibTeX{{\rm B\kern-.05em{\sc i\kern-.025em b}\kern-.08em
    T\kern-.1667em\lower.7ex\hbox{E}\kern-.125emX}}
\begin{document}

\title{Deep Learning-based Implicit CSI Feedback in Massive MIMO\\
\thanks{M. Chen, J. Guo, and S. Jin are with the National Mobile Communications Research Laboratory, Southeast University, Nanjing 210096, P. R. China (email: \{muhanchen, jiajiaguo, jinshi\}@seu.edu.cn).}
\thanks{
C.-K. Wen is with the Institute of Communications Engineering, National Sun Yat-sen University, Kaohsiung 80424, Taiwan (email: chaokai.wen@mail.nsysu.edu.tw).
}
\thanks{
G. Y. Li is with the Department of Electrical and Electronic Engineering, Imperial College London, London SW7 2AZ, U.K. (email: geoffrey.li@imperial.ac.uk).
}
\thanks{
A. Yang is with the vivo Communication Research Institute, Beijing 100020, P. R. China (ang.yang@vivo.com).
}}

\author{Muhan Chen, Jiajia Guo, \IEEEmembership{Graduate Student Member, IEEE}, Chao-Kai Wen, \IEEEmembership{Senior Member, IEEE}, Shi Jin, \IEEEmembership{Senior Member, IEEE}, Geoffrey Ye Li, \IEEEmembership{Fellow, IEEE}, and Ang Yang\vspace{-2.0em}}

\maketitle
\begin{abstract}
Massive multiple-input multiple-output can obtain more performance gain by exploiting the downlink channel state information (CSI) at the base station (BS). Therefore, studying CSI feedback with limited communication resources in frequency-division duplexing systems is of great importance. Recently, deep learning (DL)-based CSI feedback has shown considerable potential. However, the existing DL-based explicit feedback schemes are difficult to deploy because current fifth-generation mobile communication protocols and systems are designed based on an implicit feedback mechanism. In this paper, we propose a DL-based implicit feedback architecture to inherit the low-overhead characteristic, which uses neural networks (NNs) to replace the precoding matrix indicator (PMI) encoding and decoding modules. By using environment information, the NNs can achieve a more refined mapping between the precoding matrix and the PMI compared with codebooks. The correlation between subbands is also used to further improve the feedback performance. Simulation results show that, for a single resource block (RB), the proposed architecture can save $25.0\%$ and $40.0\%$ of overhead compared with Type I codebook under two antenna configurations, respectively. For a wideband system with 52 RBs, overhead can be saved by $30.7\%$ and $48.0\%$ compared with Type II codebook when ignoring and considering extracting subband correlation, respectively. 
\end{abstract}


\clearpage

\section{Introduction}
Massive multiple-input multiple-output (MIMO) technology can dramatically improve link capacity and spectrum efficiency by exploiting a large number of antennas at the base station (BS), which has been regarded as a critical technology for fifth-generation (5G) and future mobile communication systems \cite{marzetta2015massive,wong2017key,yang20196G}. However, the potential benefits provided by massive MIMO can be achieved only when the BS is with downlink channel state information (CSI). The performance gain is significantly affected by the accuracy of obtained CSI \cite{larsson2014massive}. In time-division duplexing systems, the downlink CSI can be inferred from the estimated uplink CSI at the BS by the channel reciprocity. However, no reciprocity occurs between the uplink and downlink instantaneous channels in frequency-division duplexing systems. Consequently, the downlink CSI needs to be estimated at the user equipment (UE) and then transmitted to the BS through feedback links. The feedback overhead is huge due to the large-scale transmit antennas, which consumes precious bandwidth. Therefore, studying the bandwidth-efficient CSI feedback technology is of great significance. 

The existing CSI feedback schemes are mainly based on codebook \cite{Love2008overview}, compressive sensing (CS) \cite{Kuo2012CS,Rao2014CS}, and deep learning (DL) \cite{Chen2020overview}. The commonly used codebook-based scheme in the current communication systems is designed combined with an implicit feedback mechanism \cite{3GPP2010implicit}, where the UE quantifies the partial channel information, e.g., precoding matrix indicator (PMI) and channel quality indicator, for feedback. Then, the BS reconstructs the precoding matrix based on the same designed codebook for downlink transmission. The performance of such a scheme depends on the codebook. The complexity of codebook design and the corresponding feedback overhead will increase significantly with the numbers of antennas and subcarriers. Therefore, massive MIMO brings considerable challenges to codebook-based feedback schemes. From the experimental studies in \cite{Zhou2007representation,Kyritsi2003correlation}, the channel matrices in the spatial-frequency domain tend to be sparse as the transmit antennas increase due to the limited scatters at the BS. To exploit the potential sparsity of CSI in the feedback process, the CS technique is introduced as an effective means. The channel matrices, transformed into a sparse domain, can be compressed into low-dimensional codewords for feedback by random projection and then can be reconstructed by CS algorithms, to reduce the feedback overhead. However, CS algorithms rely heavily on a prior assumption of the channel structure and suffer from high computational complexity and time expenditure, limiting the application of CS-based feedback schemes in actual systems. 

With the recent rapid development of DL technology, intelligent communication has been regarded as one of the mainstream for Beyond 5G and sixth-generation \cite{yang20196G,Hua20215Ge}. Considerable success has been achieved in wireless physical layer \cite{Wang2017deep,Qin2019deep,He2019model}, such as channel coding \cite{Nachmani2018coding}, channel estimation \cite{Soltani2019estimation,Hu2021estimation}, signal detection \cite{Ye2018detection,He2020detection}, channel prediction \cite{Yang2019prediction,Yang2020prediction}, and end-to-end transceiver design \cite{Ye2020endtoend}. As a result, many DL-based CSI feedback schemes have been developed. An autoencoder-based CSI feedback architecture, i.e., CsiNet \cite{Wen2018CsiNet}, uses neural networks (NNs) to learn how to effectively compress and reconstruct CSI. Such DL-based feedback architecture outperforms CS algorithms in terms of recovery accuracy and meets the real-time requirements of communication systems. Based on CsiNet, long short-term memory (LSTM) network is introduced into the decoder to make full use of the time correlation extracted from channels for performance improvement \cite{Wang2019CsiNetLSTM}. A CSI feedback architecture with a quantization module is designed in \cite{Guo2020CsiNet+} and an offset network is introduced to compensate for quantization distortion. Considering the interference and nonlinear effect of feedback links in practice, the denoise network in \cite{Ye2020denoise} weakens the influence of feedback noise and improves feedback performance. Network pruning and quantization techniques are introduced into the existing CSI feedback NNs in \cite{Guo2020compression} to reduce parameters, laying a foundation for practical deployment in the future. Several techniques in computer vision are applied to feedback NNs, e.g., increasing the receptive field \cite{Guo2020CsiNet+}, depth-wise separable convolution \cite{Li2020depth}, and joint convolutional residual network \cite{Lu2020bit}, to further improve the feedback accuracy. 

All the aforementioned DL-based CSI feedback architectures are designed for full channel information, which can be regarded as a scheme combining DL and explicit feedback \cite{3GPP2010implicit}. These efforts enhance CSI feedback performance, reflecting the benefits of introducing DL technology. However, the above DL-based explicit feedback brings big changes to the current air-interface standard and poses great challenges to practical deployment. Given that how to feed back and use full channel information are not systematically described and standardized in the existing 3GPP protocols, the implicit feedback mechanism is hard to be replaced with explicit feedback within a short period in the future. Therefore, we aim at retaining the framework of an implicit feedback mechanism with low-overhead characteristics and introducing new techniques to enhance the feedback performance. 

In 3GPP Release-15, Type I and Type II codebooks \cite{3GPP2019codebook} have been specified for implicit feedback in 5G new radio (NR), but they still have several problems. Type I and Type II codebooks are designed based on discrete Fourier transform (DFT) vectors \cite{Samsung2017codebooks}, which is under the premise that antenna arrays are evenly arranged in horizontal and vertical dimensions and limits antenna design and optimization. Inspired by the successful applications of DL in explicit feedback, for specific antennas and complex scenarios where Type I and Type II codebooks are not applicable, DL can use simulated or measured data to train matched NNs. DL can take full advantage of environment information to improve the feedback performance, which is difficult to achieve in codebook design. Furthermore, the channel vector in the angular domain possesses sparsity \cite{Wen2015Channel}, which is one of the characteristics used in DL-based explicit feedback. The eigenvector with direction is the partial channel information used in the existing implicit feedback mechanism. The distribution in the angular domain has certain statistical characteristics under specific scenarios. Therefore, compressibility exists in the eigenvectors. We introduce spectral entropy as the compressibility metric \cite{Sangdeh2020PSE}. Given the advantages of DL in CS, DL technology is expected to bring a breakthrough in implicit feedback. 

In this paper, we design a DL-based implicit CSI feedback architecture, named ImCsiNet, which uses NNs to replace the PMI encoding module at the UE and the PMI decoding module at the BS. Specifically, on the basis of the existing implicit feedback mechanism, the UE performs singular value decomposition (SVD) on the estimated downlink CSI and extracts the eigenvectors to form a precoding matrix. Then, instead of searching the predefined codebook to obtain the index, the precoding matrix is sent to the NNs to generate the PMI for feedback. Correspondingly, the BS uses NNs rather than the predefined codebook to reconstruct the precoding matrix from the received PMI. The major contributions of this paper are summarized as follows:

\begin{itemize}
\item We propose a novel DL-based implicit feedback architecture, which aims to retain the framework of implicit feedback mechanism in the existing standards and incorporate DL technology to enhance implicit feedback. 
\item We use NNs instead of codebooks to realize the mapping between the precoding matrix and the PMI. By making full use of environment information, the NNs can improve the feedback performance compared with Type I and Type II codebooks. 
\item We further introduce a bidirectional LSTM (bi-LSTM) network to replace the FC layers as the PMI encoder at the UE. The bi-LSTM can extract the correlation between subbands to achieve more efficient compression. Compared with FC layers, such structure is with fewer parameters and significantly reduces the feedback overhead. 
\end{itemize}

The remainder of this paper is organized as follows. Section \ref{SystemModel} introduces the massive MIMO system model and the implicit feedback scheme based on Type I and Type II codebooks in 5G NR. Section \ref{architecture} presents the DL-based implicit feedback architectures for typical massive MIMO systems. Section \ref{results} provides the experiment details and numerical results of the proposed architectures compared with Type I and Type II codebooks. Section \ref{Conclusion} concludes our work.

\section{System Model and Implicit Feedback}
\label{SystemModel}
In this section, we first introduce the typical massive MIMO systems and the SVD-based precoding algorithm. Then, we illustrate the codebook-based implicit feedback scheme in 5G NR, which serves as a performance benchmark of our design.

\subsection{Single RB}
\label{Narrowband}
Following the 5G NR standard, we consider a cyclic-prefix orthogonal frequency division multiplexing (CP-OFDM) system with a single resource block (RB), which is 180\,kHz wide in frequency and includes 12 consecutive subcarriers. The BS is equipped with $N_t$ transmit antennas and the UE is equipped with $N_r$ receiver antennas. Each antenna port at the BS occupies one subcarrier in each RB when sending a CSI-reference signal for channel measurement. Assuming that the downlink CSI can be estimated by the UE, the channel matrix on the sampled subcarrier can be defined as $\mathbf{H}\in\mathbb{C}^{N_r \times N_t}$. The element of $\mathbf{H}$, denoted as $h_{ij}$, represents the channel response between the $j$-th transmit antenna and the $i$-th receiver antenna. 

To assist the receiver to eliminate the influence of adjacent channels, precoding has been used to pre-process the transmit signals, thereby reducing the computational complexity of the receiver and maximizing the system capacity \cite{Telatar1999capacity}. SVD-based precoding is often used in massive MIMO systems. SVD of $\mathbf{H}$ is denoted as
\begin{equation}\label{SVD}
\mathbf{H} = \mathbf{U}\mathbf{\Sigma}\mathbf{V}^H,
\end{equation}
where $\mathbf{U}\in\mathbb{C}^{N_r \times N_r}$ and $\mathbf{V}\in\mathbb{C}^{N_t \times N_t}$ are the unitary matrices\footnote{The unitary matrix $\mathbf{U}$ satisfies $\mathbf{U}^H\mathbf{U}=\mathbf{U}\mathbf{U}^H=\mathbf{I}_{N_r}$ and the unitary matrix $\mathbf{V}$ satisfies $\mathbf{V}^H\mathbf{V}=\mathbf{V}\mathbf{V}^H=\mathbf{I}_{N_t}$, where $\mathbf{I}_{N_r}$ is the identity matrix of $N_r \times N_r$ and  $\mathbf{I}_{N_t}$ is the identity matrix of $N_t \times N_t$.}, $\mathbf{\Sigma}\in\mathbb{R}^{N_r \times N_t}$ is a diagonal matrix in which the diagonal elements are non-negative real numbers and the non-diagonal elements are zero. The diagonal elements of $\mathbf{\Sigma}$ are the singular values of $\mathbf{H}$, denoted as $\{\sigma_{i}, i=1,\ldots,N_{\rm{min}}\}$, where $N_{\rm{min}}=\min(N_r,N_t)$, and the singular values are sorted by default in descending order. When the downlink CSI is available at the BS, the data symbol to be transmitted can be precoded with $\mathbf{V}$ at the transmitter and then the received signal can be processed with $\mathbf{U}^H$ at the receiver. According to the properties of unitary matrices, the MIMO channel is equivalent to $N_{\rm{min}}$ parallel single-input single-output channels after processing \cite{Telatar1999capacity}. With proper power allocation, channel capacity can be maximized for a given transmit power \cite{Raleigh1998coding}, \cite{Andersen2000gain}. 

In the SVD-based precoding algorithm, $\mathbf{V}$ plays an important role in the system performance and needs to be fed back to serve as the partial channel information. Therefore, in the implicit feedback scheme, how to design codebooks for an accurate CSI representation is the core of this scheme. The main purpose of this paper is to integrate DL into the existing implicit feedback to achieve more efficient feedback, which provides a new direction for designing CSI feedback. To simplify the descriptions, we only consider the case of single-stream transmission in this paper. Therefore, the precoding vector to be fed back is the eigenvector corresponding to the maximum singular value in $\mathbf{V}$ calculated in (\ref{SVD}), which is denoted as $\mathbf{v}\in\mathbb{C}^{N_t \times 1}$. In Section \ref{architecture}, we design NNs to implement the feedback of eigenvector $\mathbf{v}$. The proposed DL-based implicit feedback architecture can be easily extended to multi-stream transmission by changing the structure of NNs to multi-channel convolutional neural networks.

\subsection{Multiple RBs}
\label{Wideband}
We extend the process of extracting eigenvectors for generating precoding matrices to the CP-OFDM system with $N_{\rm{RB}}$ RBs. The entire band is divided into $N_s$ subbands and each subband contains $N_{\rm{RB}}/N_{s}$ RBs. The channel matrix measured on the $n$-th RB of the $s$-th subband can be denoted as $\mathbf{H}_{s,n}\in\mathbb{C}^{N_r \times N_t}$, for $n=1,\ldots,N_{\rm{RB}}/N_{s}$ and $s=1,\ldots,N_{s}$. Given the SVD of $\mathbf{H}_{s,n}$ and combined with the properties of unitary matrices, the following eigenvalue decomposition (EVD) is established: 
\begin{equation}\label{EVD}
\mathbf{H}_{s,n}^H\mathbf{H}_{s,n} = \mathbf{V}_{s,n}\mathbf{\Lambda}_{s,n}\mathbf{V}_{s,n}^H,
\end{equation}
where $\mathbf{H}_{s,n}^H\mathbf{H}_{s,n}\in\mathbb{C}^{N_t \times N_t}$ is a Hermitian matrix, $\mathbf{V}_{s,n}\in\mathbb{C}^{N_t \times N_t}$ is the unitary matrix same as that obtained by the SVD of $\mathbf{H}_{s,n}$, $\mathbf{\Lambda}_{s,n}=\mathbf{\Sigma}_{s,n}^H\mathbf{\Sigma}_{s,n}\in\mathbb{R}^{N_t \times N_t}$ is a diagonal matrix, and the diagonal elements are the square of the singular values $\{\sigma_i\}$.

Subband feedback is supported in 5G NR \cite{3GPP2019codebook}. Therefore, we extract the eigenvectors of each subband as the channel information to be sent back. Specifically, we first calculate the value of $\mathbf{H}_{s,n}^H\mathbf{H}_{s,n}$ for each RB. The values of all RBs in one subband are averaged to represent the subband characteristics. Then, EVD is applied to the average $\mathbf{H}_{s,n}^H\mathbf{H}_{s,n}$ for each subband. We denote the unitary matrix obtained by EVD as $\mathbf{V}_{s}\in\mathbb{C}^{N_t \times N_t}$, where $s$ is the subband index. The eigenvector corresponding to the maximum eigenvalue in $\mathbf{V}_{s}$, denoted as $\mathbf{v}_{s}\in\mathbb{C}^{N_t \times 1}$, serves as the channel information that needs to be fed back for the $s$-th subband. Finally, the eigenvectors extracted from $N_s$ subbands are stacked to form the eigen matrix over the entire band, which can be written as
\begin{equation}\label{Vstack}
\mathbf{V}_{\rm{stack}} = [\mathbf{v}_{1}, \mathbf{v}_{2}, \ldots, \mathbf{v}_{N_{s}}].
\end{equation}
where $\mathbf{V}_{\rm{stack}}\in\mathbb{C}^{N_t \times N_s}$ and the original overhead is $N_{s}$ times that of an eigenvector in the case of single RB. 

Given that channel matrices corresponding to the adjacent subcarriers are correlated, their eigenvectors corresponding to the maximum eigen values, $\mathbf{v}_{1}, \ldots, \mathbf{v}_{N_{s}}$, are correlated. Even if the correlation can be effectively described in the Grassman manifold, it is hard to be directly used for compression. Fortunately, the issue can be resolved with DL. We design suitable NNs to compress eigen matrix $\mathbf{V}_{\rm{stack}}$ for feedback in Section \ref{architecture}.

\begin{figure*}[tb]
    \centering
     \subfigure[A single-panel antenna]{
     \label{antenna}
     \includegraphics[width=0.43\textwidth,trim=10 8 13 12,clip]{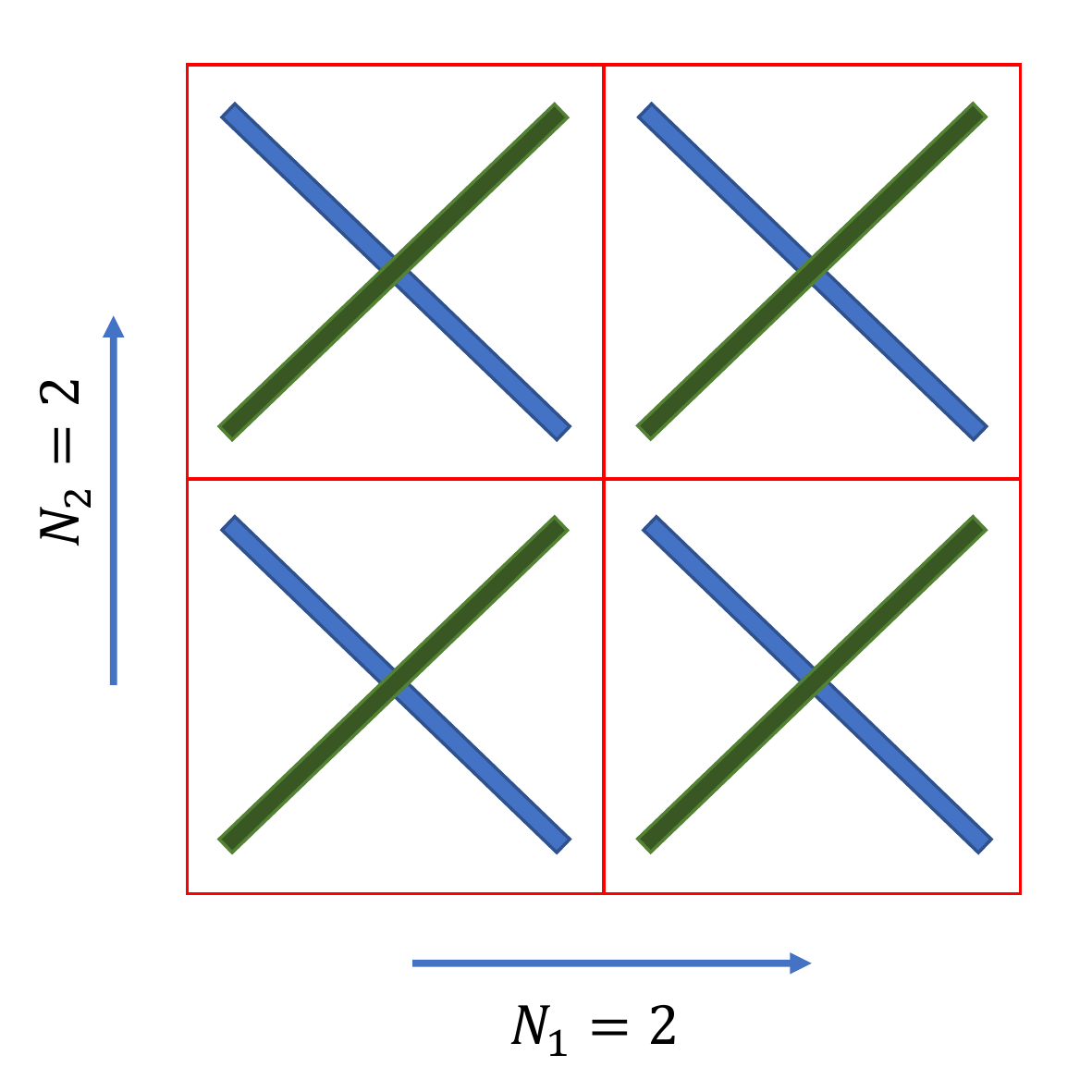} 
     }
    \quad
     \subfigure[2D DFT based gird of beams]{
     \label{beam}
     \includegraphics[width=0.5\textwidth,trim=12 20 14 12,clip]{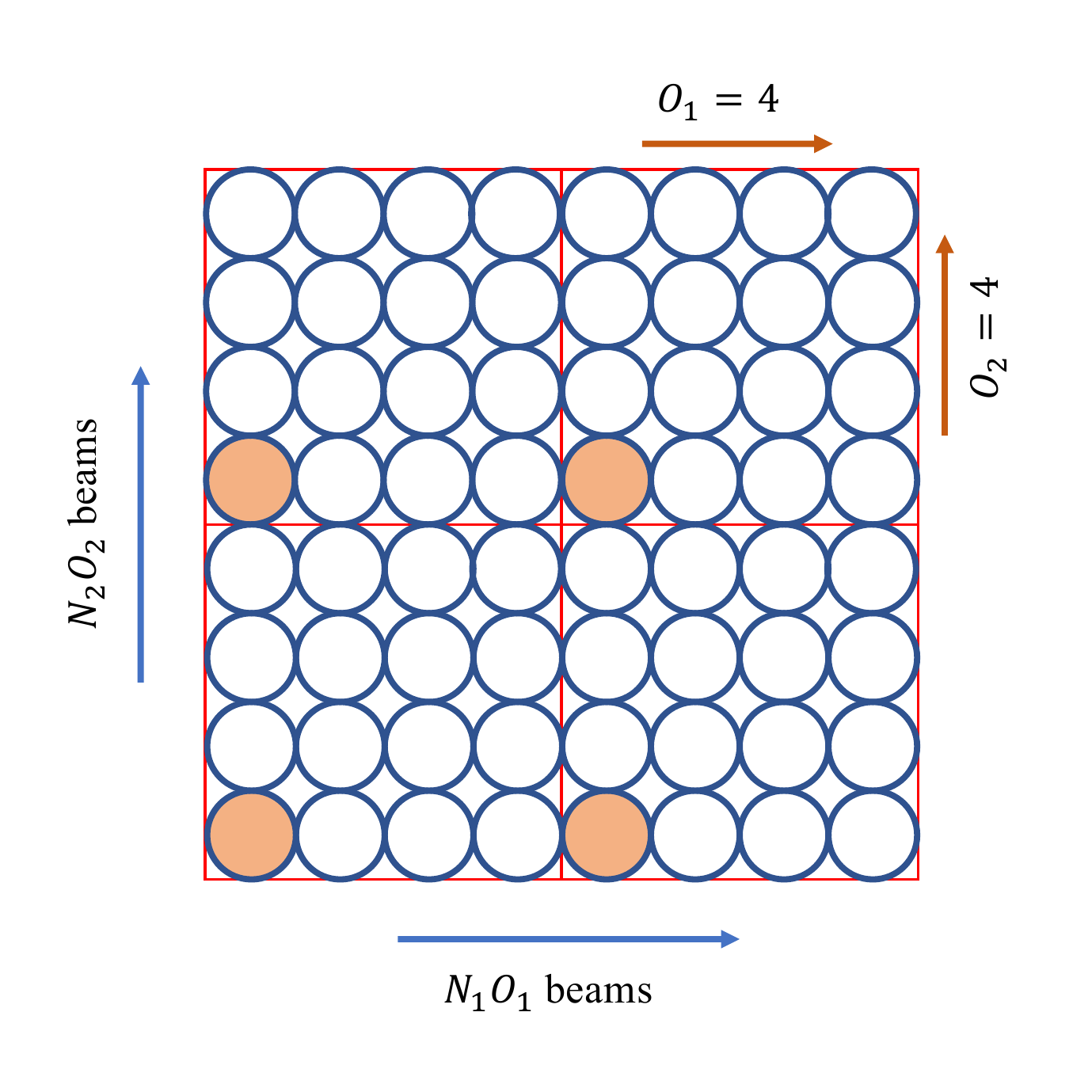} 
     }
	\caption{(a) Example of a cross-polarized single-panel antenna array with $(N_1,N_2)=(2,2)$; (b) 2D oversampled DFT beams with $(O_1,O_2)=(4,4)$.} 
	\label{fig:subfig}
\end{figure*}

\subsection{Codebook-based CSI Feedback in 5G NR}
\label{codebook}
In the 5G NR standard, two types of codebooks are supported, i.e., Type I codebook with low resolution and Type II codebook with high resolution. They are designed based on DFT beams and follow the design principle of dual-stage codebook structure in LTE \cite{Yun2011double}. Fig. \ref{antenna} is a schematic diagram of a single-panel antenna array configured with $(N_1, N_2)$, where $N_{1}$ and $N_{2}$ represent the numbers of antenna ports in the horizontal and vertical dimensions in the same polarization direction, respectively. The total number of transmit antenna ports is $N_{t}=2N_{1}N_{2}$. To refine the beam granularity for an accurate direction, oversampling factors $(O_1, O_2)$ are introduced to form a two-dimensional (2D) DFT-based grid of beams \cite{Ericsson2016structure}, as shown in Fig. \ref{beam}. The one-dimensional oversampled DFT beams in the horizontal and vertical dimensions can be respectively represented as
\begin{equation}\label{u_h}
\bm{\mu}^{[\rm h]} = [1 \quad e^{j\frac{2\pi\theta_{1}}{N_{1}O_{1}}} \quad \ldots \quad e^{j\frac{2\pi\theta_{1}(N_{1}-1)}{N_{1}O_{1}}}]^T, \quad \theta_{1}=0,1,\ldots,N_{1}O_{1}-1,
\end{equation}
\begin{equation}\label{u_h}
\bm{\mu}^{[\rm v]} = [1 \quad e^{j\frac{2\pi\theta_{2}}{N_{2}O_{2}}} \quad \ldots \quad e^{j\frac{2\pi\theta_{2}(N_{2}-1)}{N_{2}O_{2}}}]^T, \quad \theta_{2}=0,1,\ldots,N_{2}O_{2}-1,
\end{equation}
where $\bm{\mu}^{[\rm h]}\in\mathbb{C}^{N_{1}\times1}$, $\bm{\mu}^{[\rm v]}\in\mathbb{C}^{N_{2}\times1}$, $\theta_{1}$ and $\theta_{2}$ refer to the beam index in the horizontal and the vertical dimensions, respectively.
Given the antenna port and the oversampling factor settings according to \cite{3GPP2019codebook}, a set of 2D DFT beams in the spatial domain can be obtained as
\begin{equation}\label{2Dbeam}
\mathcal{D} = \left\{\bm{b}_{\rm{\theta_{1}},\rm{\theta_{2}}}|\bm{b}_{\rm{\theta_{1}},\rm{\theta_{2}}} = \bm{\mu}^{[\rm h]} \otimes \bm{\mu}^{[\rm v]}\right\},
\end{equation}
where $\otimes$ represents Kronecker product, $\bm{b}_{\rm{\theta_{1}},\rm{\theta_{2}}}\in\mathbb{C}^{N_{1}N_{2}\times1}$ is an oversampled 2D DFT beam and $\mathcal{D}$ contains a total of $N_{1}O_{1} \times N_{2}O_{2}$ beams.

With the 2D DFT beam set $\mathcal{D}$, Type I codebook is generated based on beam selection and phase adjustment \cite{Intel2017TypeI}, and Type II codebook is generated by selecting a beam subset and choosing linear combination coefficients for beam merging \cite{Samsung2017TypeII}. The same beam subset is used for different polarization directions. For rank 1 (single-stream transmission), Type I codebook can be defined as
\begin{equation}\label{Type I}
\mathbf{w}_{\rm{I}} = \frac{1}{\sqrt{2N_{1}N_{2}}}
\begin{bmatrix}
\bm{b}_{\rm{\theta_{1}},\rm{\theta_{2}}} \\
\varphi\bm{b}_{\rm{\theta_{1}},\rm{\theta_{2}}} \\
\end{bmatrix},
\end{equation}
where $\varphi\in\left\{1,j,-1,-j\right\}$ is used to quantify the phase difference between the two polarization directions. $\mathbf{w}_{\rm{I}}\in\mathbb{C}^{2N_{1}N_{2}\times1}$ has the same dimension as eigenvector $\mathbf{v}$. The number of $\mathbf{w}_{\rm{I}}$ in the codebook set is $4N_{1}O_{1}N_{2}O_{2}$, hence the bit number required by PMI is $\log_2(4N_{1}O_{1}N_{2}O_{2})$. 

For rank 1, Type II codebook for one subband can be expressed as
\begin{equation}\label{Type II}
\mathbf{w}_{\rm{II}} = 
\begin{bmatrix}
\sum_{i=0}^{K}\bm{b}_{\theta_{1}^{(i)},\theta_{2}^{(i)}}p_{0,i}^{\rm{WB}}p_{0,i}^{\rm{SB}}c_{0,i} \\
\sum_{i=0}^{K}\bm{b}_{\theta_{1}^{(i)},\theta_{2}^{(i)}}p_{1,i}^{\rm{WB}}p_{1,i}^{\rm{SB}}c_{1,i} \\
\end{bmatrix},
\end{equation}
where $\mathbf{w}_{\rm{II}}\in\mathbb{C}^{2N_{1}N_{2}\times1}$ is with the same dimension as subband eigenvector $\mathbf{v}_{s}$, each row represents a linear combination of $K$ beams selected from $\mathcal{D}$ in one polarization direction; $p_{r,i}^{\rm{WB}}$ and $p_{r,i}^{\rm{SB}}$ stand for the wideband amplitude combining coefficient and the subband amplitude combining coefficient for beam $i$ and on polarization $r$, respectively; $c_{r,i}$ denotes the phase combining coefficient for beam $i$ and on polarization $r$, which can be configured as QPSK or 8PSK. The amplitude and phase combining coefficients of each subband are calculated separately. 

For $N_{s}$ subbands, Type II codebook over the entire band can be denoted as $\mathbf{W}_{\rm{II}}\in\mathbb{C}^{2N_{1}N_{2} \times N_{s}}$, which corresponds to eigen matrix $\mathbf{V}_{\rm{stack}}$. The feedback overhead of Type II codebook is variable, which is related to the number of zero elements in wideband amplitude combining coefficient $p_{r,i}^{\rm{WB}}$ \cite{3GPP2019codebook}. Therefore, we use the average overhead of all test samples to indicate the feedback overhead of Type II codebook in this paper. On the basis of Type I and Type II codebooks, the UE selects the most similar codebook for feedback according to the measured channel information. In the subsequent experiments, we test the performance of the two codebook-based methods as a baseline for the DL-based implicit feedback.

\section{DL-based Implicit Feedback Architecture}
\label{architecture}
In this section, we introduce the DL-based implicit feedback architecture. First, a basic framework composed of FC layers is proposed for the CP-OFDM system with a single RB. Then, it is extended to multiple RBs. Bidirectional LSTMs are introduced to extract the correlation between subbands for performance improvement. 

\subsection{DL-based Feedback for Single RB}
The core of implicit feedback design is the trade-off between the channel representation resolution and the feedback overhead. To reduce the feedback overhead without causing significant performance loss, an effective means for dimensionality reduction needs to be introduced. Inspired by DL-based explicit feedback normally following the autoencoder framework for compression, we design a DL-based implicit feedback architecture, which uses NNs to replace the PMI encoding module (encoder) at the UE and the PMI decoding module (decoder) at the BS, respectively. The distribution of channels in the angular domain has certain statistical characteristics under specific scenarios. By using the environment information, the NNs can effectively learn a low-dimensional representation for precoding matrices, i.e., PMI. In order to quantify the compressibility of the eigenvectors, we introduce power spectral entropy (PSE) \cite{Sangdeh2020PSE} as the compressibility metric. For an eigenvector $\mathbf{v}=[v_{1}, v_{2}, \ldots, v_{N_{t}}]$, we denote its DFT output as $\bm{\vartheta}=[\vartheta_{1}, \vartheta_{2}, \ldots, \vartheta_{N_{t}}]$. The PSE of $\mathbf{v}$ can be defined as
\begin{equation}
\label{PSE}
{\textrm{PSE}}(\mathbf{v}) = -\frac{1}{{\rm{log_2}}N_t}\sum_{i=1}^{N_t}p(\vartheta_{i}){\rm{log_2}}p(\vartheta_{i}),
\end{equation}
where $p(\vartheta_{i})=\frac{|\vartheta_{i}|^2}{\sum_{i=1}^{N_t}|\vartheta_{i}|^2}$ \cite{Heurtier2016SE}. The PSE value is normalized to $[0, 1]$. A low PSE value means that energy in the angular domain is relatively concentrated and indicates high compressibility. 

\begin{figure*}[tb]
    \centering 
     \includegraphics[width=\textwidth]{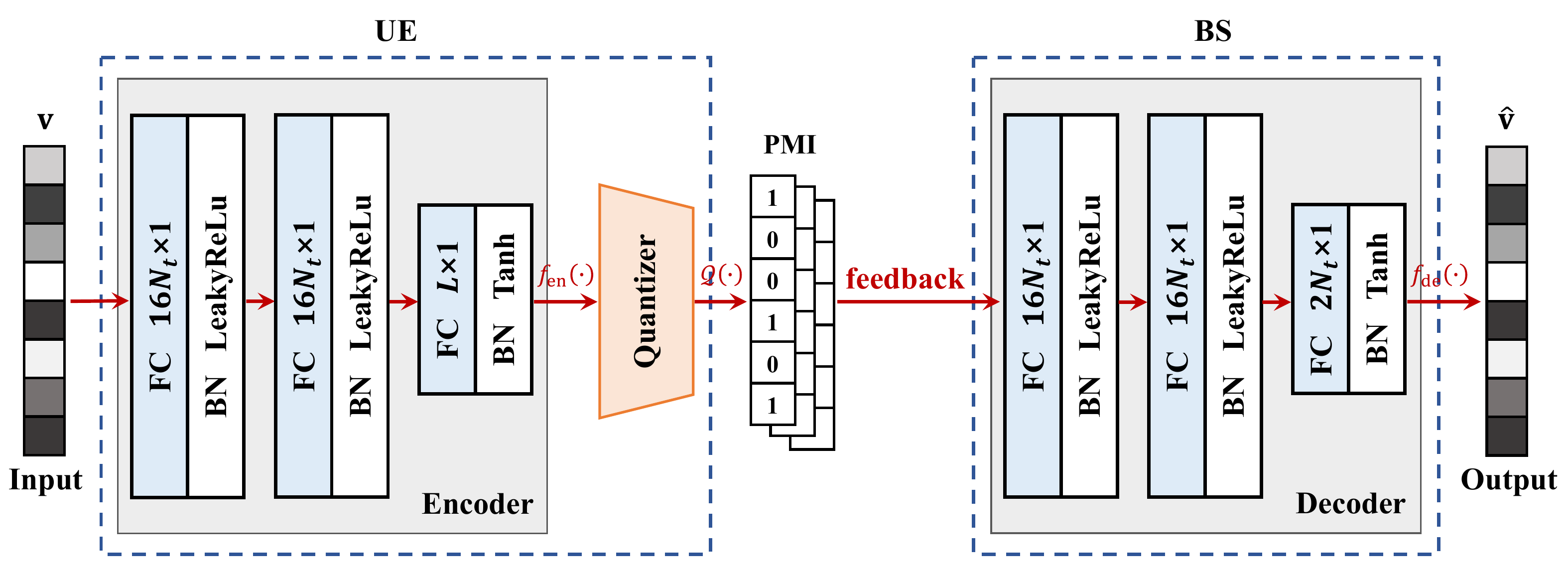}    
	\caption{\label{ImCsiNet-s}Architecture of ImCsiNet-s for single RB. NNs are used to replace the PMI encoding module at the UE and the PMI decoding module at the BS. The input of the encoder is the eigenvector $\mathbf{v}$ extracted from the full channel matrix. } 
\end{figure*}

On the basis of the measured compressibility, the autoencoder framework compresses and reconstructs the eigenvectors as in Fig. \ref{ImCsiNet-s}. Since the practical transmission is with a finite number of bits, a quantization module is added after the encoder to generate a bitstream. The feedback overhead is determined by the output dimension of the encoder and the number of quantization bits at the quantizer. When the output dimension of the encoder is $L$ and the quantizer uses $B$-bit quantization for each element of the compressed codeword, the total number of feedback bits can be expressed as
\begin{equation}
\label{feedback_bits}
N_{\rm{bits}} = L \times B.
\end{equation}
Different quantization methods can be used to improve feedback performance based on datasets. 
In this paper, binarization operation is used for quantization for single RB due to the low dimension of feedback information, which is equivalent to $B=1$. For multiple RBs, $B$-bit uniform quantization is adopted since the number of feedback bits is large. The feedback overhead is discussed in detail together with quantizers subsequently. The basic framework contains three modules, i.e., encoder, quantizer, and decoder, which is called ImCsiNet. For single-RB and multi-RB systems, appropriate network layers and quantizers are filled on the basis of ImCsiNet.

For the CP-OFDM system with a single RB, the input of the encoder is the eigenvector $\mathbf{v}$ extracted from the full channel matrix. The eigenvector is encoded to generate compressed codewords and then quantized to obtain the corresponding PMI for feedback. The decoder at the BS recovers the eigenvector from the PMI. The entire process for DL-based implicit feedback can be formulated as
\begin{equation}
\label{autoencoder}
\hat{\mathbf{v}} = f_{\rm{de}}(\mathcal{Q}(f_{\rm{en}}(\mathbf{v},\Theta_{\rm{en}})),\Theta_{\rm{de}}),
\end{equation}
where $f_{\rm{en}}(\cdot)$ and $f_{\rm{de}}(\cdot)$ represent the operations of the encoder and the decoder, respectively; the quantization operation, $\mathcal{Q}(\cdot)$, converts the compressed codewords into bitstreams, which also introduces quantization errors; $\Theta_{\rm{en}}$ and $\Theta_{\rm{de}}$ are the parameters of the encoder and the decoder, respectively. 

Given that the feedback information is the eigenvector with direction, the purpose of model training is to recover the direction information of the eigenvector to the largest extent. Therefore, the cosine similarity in \cite{Wen2018CsiNet} is used to evaluate the reconstruction performance of both the codebook-based and DL-based implicit feedback methods. The cosine similarity, denoted as $\rho_{\rm{s}}$, can be given by
\begin{equation}
\label{rho_s}
\rho_{\rm{s}} = \frac{\left|\hat{\mathbf{v}}^{H}\mathbf{v}\right|}{\left\|\hat{\mathbf{v}}\right\|_{2}\left\|\mathbf{v}\right\|_{2}},
\end{equation}
where $\left\|\cdot\right\|_{2}$ is the Euclidean norm. The range of $\rho_{\rm{s}}$ is obviously $[0,1]$ from the Cauchy-Schwarz inequality. The closer the $\rho_{\rm{s}}$ is to 1, the more similar the reconstructed $\hat{\mathbf{v}}$ is to the original $\mathbf{v}$. Accordingly, we take the negative of the cosine similarity as the loss function, which can be written as
\begin{equation}
\label{loss_s}
{\cal L}_{\rm{s}} = -\frac{1}{N}\sum_{i=1}^{N}\frac{\left|\hat{\mathbf{v}}_{i}^{H}\mathbf{v}_{i}\right|}{\left\|\hat{\mathbf{v}}_{i}\right\|_{2}\left\|\mathbf{v}_{i}\right\|_{2}},
\end{equation}
where $N$ is the number of training samples. 

\begin{table*}[tb]
\centering
\caption{Details of the proposed ImCsiNet-s}
\begin{threeparttable}
\setlength{\tabcolsep}{3.5mm}{
\begin{tabular}{c|ccccc} \hline \hline
                         & Layer name & Output shape  & Activation function & Parameters & FLOPs \\
\hline \hline
\multirow{4}{*}{Encoder} & Input    & $2N_t \times 1$   & $\backslash$  & 0                         & 0 \\
                         & FC1+BN1  & $16N_t \times 1$  & LeakyReLu     & $16N_t \times (2N_t+5)$   & $16N_t \times (4N_t-1)$ \\
                         & FC2+BN2  & $16N_t \times 1$  & LeakyReLu     & $16N_t \times (16N_t+5)$  & $16N_t \times (32N_t-1)$ \\
                         & FC3+BN3  & $L \times 1$      & Tanh          & $L \times (16N_t+5)$      & $L \times (32N_t-1)$ \\ 
\hline \hline
\multirow{3}{*}{Decoder} & FC4+BN4  & $16N_t \times 1$  & LeakyReLu     & $16N_t \times (L+5)$      & $16N_t \times (2L-1)$ \\
                         & FC5+BN5  & $16N_t \times 1$  & LeakyReLu     & $16N_t \times (16N_t+5)$  & $16N_t \times (32N_t-1)$ \\
                         & FC6+BN6  & $2N_t \times 1$   & Tanh          & $2N_t \times (16N_t+5)$   & $2N_t \times (32N_t-1)$ \\
\hline \hline
\end{tabular}}
\label{Details of ImCsiNet-s}
\begin{tablenotes}
\footnotesize
\item[1] The quantization module without training parameters is neglected.
\item[2] The FLOPs of the activation functions can be ignored compared to the FC layers.
\end{tablenotes}
\end{threeparttable}
\end{table*}

As shown in Fig. \ref{ImCsiNet-s}, an architecture composed of FC layers, namely ImCsiNet-s, is designed for single RB. The first two FC layers in the encoder are used to extract the channel features, and the last FC layer with $L$ neurons is used to compress the channel data into the compressed codeword with a dimension of $L \times 1$. Then, the codeword is quantized by a quantizer to generate PMI with finite bits for feedback. Once the BS receives the PMI, the FC layers in the decoder are used for decompression and reconstruction to obtain the eigenvector corresponding to the PMI. When the number of feedback bits is very small, binarization operation outperforms uniform quantization \cite{Guo2020cooperative}. Therefore, binarization operation is adopted in the quantization module of ImCsiNet-s. For the normalized value of the compressed codeword, denoted as $x\in[-1,1]$, binarization operation generates a discrete output valued in the set $\{-1,1\}$, which is defined as
\begin{equation}
\label{binarization}
\begin{split}
&\qquad \qquad \qquad \mathcal{Q}_{\rm{b}}(x) = x + \epsilon, \\
\epsilon &= \left\{
\begin{aligned}
& 1-x,   &{\rm{with \quad probability \quad}} \frac{1+x}{2} \\
& -(1+x),  &{\rm{with \quad probability \quad}} \frac{1-x}{2}
\end{aligned}
\right.,
\end{split}
\end{equation}
where $\epsilon$ denotes the quantization noise with zero mean. According to (\ref{binarization}), we design a quantization layer that performs such a binarization operation. Even if binarization is non-differentiable, it can be regarded as a transparent operation \cite{Theis2017lossy}. Therefore, the gradient can be backpropagated through the quantization layer without loss and the gradient of the quantization layer can be set as 1 to realize the backpropagation algorithm. The binarization operation is equivalent to an optimized 1-bit quantization. Therefore, the final feedback overhead is equal to the number of neurons in the last FC layer of the encoder, i.e., $N_{\rm{bits}} = L \times 1$.

The details of ImCsiNet-s are shown in Table \ref{Details of ImCsiNet-s}, including the layer structure, parameters, and floating-point operations (FLOPs). The real and imaginary parts of the eigenvector $\mathbf{v} \in \mathbb{C}^{N_t \times 1}$ are spliced together to form a real-valued input vector. Thus, the input dimension of the encoder is $2N_{t} \times 1$. A batch normalization (BN) layer is added between all the FC layers and activation functions to alleviate gradient dispersion problems and improve the generalization performance of the network. The LeakyReLU function is used as the activation function of the first two BN layers both in the encoder and the decoder to increase the nonlinearity of the network. The activation function of the last BN layer in the encoder and the decoder is the Tanh function to normalize the output into the range of $(-1,1)$. The parameters and FLOPs are generally used to measure the network complexity. According to \cite{Molchanov2017pruning}, the parameters and FLOPs of FC layers are calculated by
\begin{equation}
\label{parameters_FC}
N_{\rm{FC}} = O \times (I+1), 
\end{equation}
\begin{equation}
\label{FLOPs_FC}
{\rm{FLOPs}}_{\rm{FC}} = O \times (2I-1),
\end{equation}
where $I$ and $O$ denote the input and the output dimensionality, respectively. The parameters of BN layers can be calculated by 
\begin{equation}
\label{parameters_BN}
N_{\rm{BN}} = 4O.
\end{equation}
Therefore, the total parameter number of the combination of one FC layer and one BN layer is
\begin{equation}
\label{parameters_FC+BN}
N_{\rm{FC+BN}} = O \times (I+5). 
\end{equation}
When the BN layer lies between the FC layer and the activation function, its parameters can be folded into the weight calculation of the FC layer by an equivalent transformation. The effect of the BN layer can be ignored when calculating FLOPs for this reason. For ease of comparison, the output dimension of the encoder can be expressed as $L=\alpha \times 2N_{t}$, where $2N_{t}$ is the dimension of the real-valued input vector, $\alpha \in (0,1)$ denotes the compression ratio. In this case, the total parameter number of the proposed ImCsiNet-s is $(576+64\alpha)N_{t}^{2}+(330+10\alpha)N_{t}$ and the total FLOPs number is $(1152+128\alpha)N_{t}^{2}-(66+2\alpha)N_{t}$.

\subsection{DL-based Feedback for Multiple RBs}
\label{architecture for wideband}
The architecture of ImCsiNet-s can be extended to multiple RBs, called ImCsiNet-m. The network structure of ImCsiNet-m is similar to that of ImCsiNet-s. The details of ImCsiNet-m are shown in Table \ref{Details of ImCsiNet-m}. The network depth of ImCsiNet-m remains the same as that of ImCsiNet-s while the network width increases due to the increasing feedback information. The feedback information is the stack of the eigenvectors extracted from $N_{s}$ subbands, i.e., $\mathbf{V}_{\rm{stack}} \in \mathbb{C}^{N_{t} \times N_{s}}$. The eigenvectors in the eigen matrix $\mathbf{V}_{\rm{stack}}$ are spliced in the vertical direction and then the real and imaginary parts are extracted and concatenated to form a real-valued vector. The real-valued vector is used as the input of ImCsiNet-m. Therefore, the input dimension of the encoder is $2N_{t}N_{s}\times1$. On this basis, the real-valued input vector is compressed by the encoder to generate a compressed codeword with a dimension of $L \times 1$, and then quantized to obtain corresponding PMI for feedback. The BS recovers the real-valued vector according to the received PMI in the decoder. Finally, the recovered vector is reshaped into the eigen matrix, denoted as $\hat{\mathbf{V}}_{\rm{stack}}$. The performance metric of the network can be given by
\begin{equation}
\label{rho_m}
\rho_{\rm{m}} = \frac{1}{N_{s}}\sum_{s=1}^{N_{s}}\frac{\left|\hat{\mathbf{v}}_{s}^{H}\mathbf{v}_{s}\right|}{\left\|\hat{\mathbf{v}}_{s}\right\|_{2}\left\|\mathbf{v}_{s}\right\|_{2}},
\end{equation}
where $\mathbf{v}_{s}$ is the column vector (eigenvector) that makes up the eigen matrix $\mathbf{V}_{\rm{stack}}$. The range of $\rho_{\rm{m}}$ is $[0,1]$. To make recovered $\hat{\mathbf{V}}_{\rm{stack}}$ as similar to original $\mathbf{V}_{\rm{stack}}$ as possible, that is, $\rho_{\rm{m}}$ as close as possible to 1, the loss function is set as
\begin{equation}
\label{loss_m}
{\cal L}_{\rm{m}} = -\frac{1}{N}\frac{1}{N_{s}}\sum_{i=1}^{N}\sum_{s=1}^{N_{s}}\frac{\left|\hat{\mathbf{v}}_{s,i}^{H}\mathbf{v}_{s,i}\right|}{\left\|\hat{\mathbf{v}}_{s,i}\right\|_{2}\left\|\mathbf{v}_{s,i}\right\|_{2}},
\end{equation}
where $N$ is the number of training samples. 

\begin{table*}[tb]
\centering
\caption{Details of the proposed ImCsiNet-m}
\setlength{\tabcolsep}{2.8mm}{
\begin{tabular}{c|ccccc} \hline \hline
                         & Layer name & Output shape  & Activation function & Parameters & FLOPs \\
\hline \hline
\multirow{4}{*}{Encoder} & Input    & $2N_{t}N_{s} \times 1$  & $\backslash$  & 0                      & 0 \\
                         & FC1+BN1  & $8N_{t}N_{s} \times 1$  & LeakyReLu     & $8N_{t}N_{s} \times (2N_{t}N_{s}+5)$  & $8N_{t}N_{s} \times (4N_{t}N_{s}-1)$ \\
                         & FC2+BN2  & $8N_{t}N_{s} \times 1$  & LeakyReLu     & $8N_{t}N_{s} \times (8N_{t}N_{s}+5)$  & $8N_{t}N_{s} \times (16N_{t}N_{s}-1)$ \\
                         & FC3+BN3  & $L \times 1$            & Tanh          & $L \times (8N_{t}N_{s}+5)$            & $L \times (16N_{t}N_{s}-1)$ \\ 
\hline \hline
\multirow{3}{*}{Decoder} & FC4+BN4  & $8N_{t}N_{s} \times 1$  & LeakyReLu     & $8N_{t}N_{s} \times (L+5)$            & $8N_{t}N_{s} \times (2L-1)$ \\
                         & FC5+BN5  & $8N_{t}N_{s} \times 1$  & LeakyReLu     & $8N_{t}N_{s} \times (8N_{t}N_{s}+5)$  & $8N_{t}N_{s} \times (16N_{t}N_{s}-1)$ \\
                         & FC6+BN6  & $2N_{t}N_{s} \times 1$  & Tanh          & $2N_{t}N_{s} \times (8N_{t}N_{s}+5)$  & $2N_{t}N_{s} \times (16N_{t}N_{s}-1)$ \\
\hline \hline
\end{tabular}}
\label{Details of ImCsiNet-m}
\end{table*}

Considering that the dimension of eigen matrix $\mathbf{V}_{\rm{stack}}$ is large, the required feedback bits are much larger than that of the single-RB system. Therefore, uniform quantization is adopted in the quantization module of ImCsiNet-m. When using $B$-bit uniform quantization to quantize the normalized value of the compressed codeword, i.e., $x\in[-1,1]$, the output can be expressed as
\begin{equation}
\label{uniformquantization}
\mathcal{Q}_{\rm{u}}(x) = \frac{{\rm round}(x\cdot2^{B-1})}{2^{B-1}}, 
\end{equation}
where ${\rm round}(\cdot)$ denotes an approximate rounding operation, which is non-differentiable. Similarly, the backpropagation algorithm can be successfully implemented by setting the gradient of the quantization layer to 1. In this case, the final feedback overhead is $N_{\rm{bits}} = L \times B$. The number of feedback bits per subband can be denoted as $N_{\rm{ps}}$. The relationship between the total feedback overhead and the subband overhead can be expressed as $N_{\rm{bits}} = N_{s} \times N_{\rm{ps}}$.

From Table \ref{Details of ImCsiNet-m}, we can calculate the network complexity of ImCsiNet-m. The output dimension can be represented by compression ratio $\alpha$ and input dimension $2N_{t}N_{s}$, i.e., $L = \alpha \times 2N_{t}N_{s}$. Thus, the total parameter number of the proposed ImCsiNet-m is $(160+32\alpha)N_{s}^{2}N_{t}^{2}+(170+10\alpha)N_{s}N_{t}$ and the total FLOPs number is $(320+64\alpha)N_{s}^{2}N_{t}^{2}-(34+2\alpha)N_{s}N_{t}$. For the wideband system with multiple RBs, the dimension of channel data to be compressed is $N_{s}$ times larger than that of single RB under the same antenna configuration, which leads to a large number of parameters in ImCsiNet-m composed of FC layers. In addition, the eigen matrix is reshaped into a vector to input the FC layers, which fails to make full use of the correlation between subbands. Therefore, adjusting the network structure is necessary to reduce parameters and achieve more efficient compression. 

\begin{figure*}[tb]
    \centering
     \subfigure[Architecture of bi-ImCsiNet]{
     \label{fig:subfig:a}
     \includegraphics[width=\textwidth]{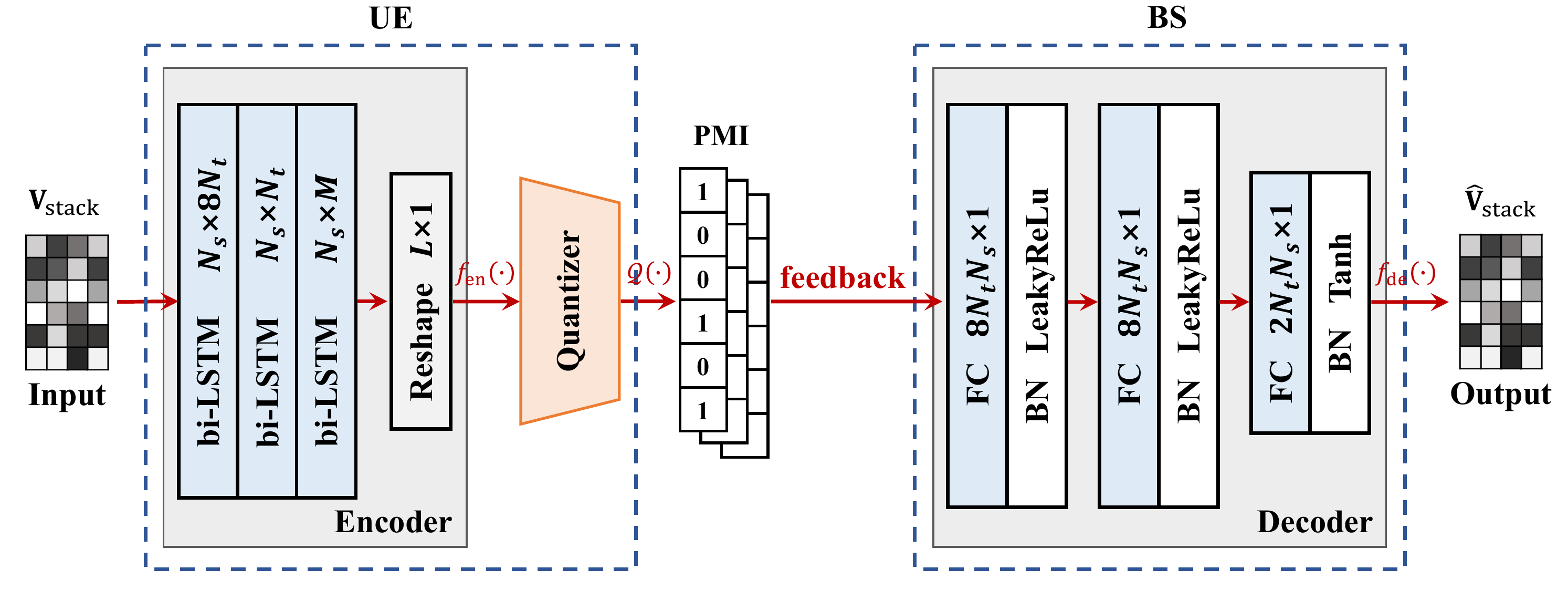} 
     }
    \quad
     \subfigure[A bi-LSTM layer]{
     \label{fig:subfig:b}
     \includegraphics[width=14cm]{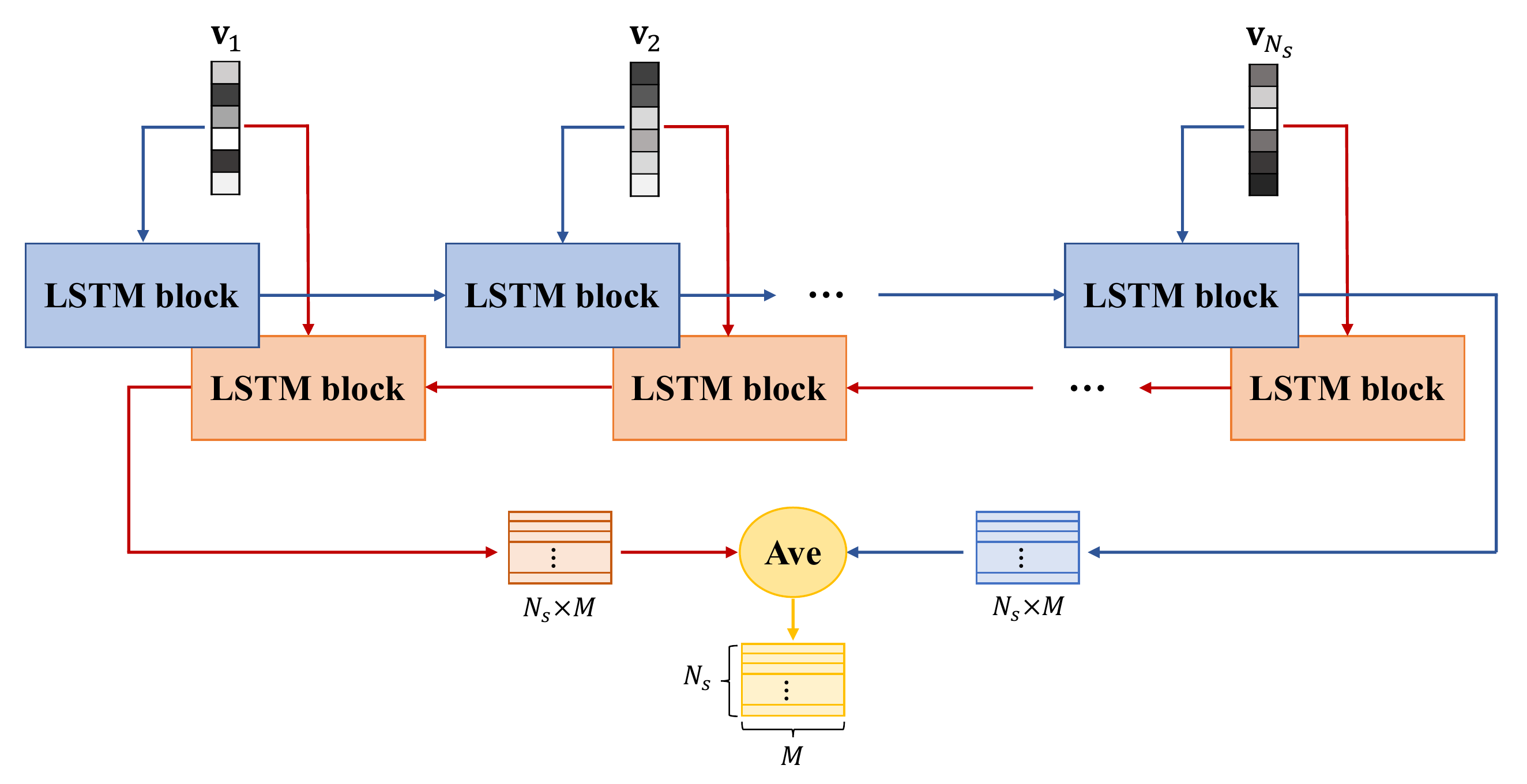} 
     }
	\caption{(a) Architecture of bi-ImCsiNet, which uses bi-LSTMs instead of FC layers to extract the subband correlation; (b) The structure of a layer in the bi-LSTM network in the bi-ImCsiNet encoder.} 
	\label{fig:subfig}
\end{figure*}

Inspired by the impressive performance of recurrent neural networks (RNNs) \cite{Elman1990LSTM} in natural language processing and extracting correlation among sequence data, we use RNNs to replace the FC layers in the encoder to extract the correlation of eigenvectors among different subbands, to reduce the feedback overhead while ensuring the reconstruction performance. A three-layer bidirectional LSTM (bi-LSTM) network is used as the encoder for sufficient extraction. The new architecture, called bi-ImCsiNet, is shown in Fig. \ref{fig:subfig:a}. As expressed in (\ref{Vstack}), $\mathbf{V}_{\rm{stack}}$ contains $N_{s}$ eigenvectors of different subbands. Different from ImCsiNet-m jointly compressing $N_{s}$ eigenvectors into an $L$-dimensional codeword, bi-ImCsiNet compresses the eigenvector of each subband into an $M$-dimensional codeword respectively, where $L = N_{s} \times M$. As a result, the time step of the bi-LSTM network is set as $N_{s}$. The LSTM blocks in one direction share parameters, thus reducing the network storage overhead. 

Fig. \ref{fig:subfig:b} shows the structure of a layer in the bi-LSTM network, which contains two LSTMs to process the sequence of eigenvectors in different orders. When the sequence structure is maintained, two compressed matrices with a dimension of $N_{s} \times M$ are obtained in the forward and backward LSTMs, and then the average of the two matrices are calculated as the final output. The details of the encoder in bi-ImCsiNet are shown in Table \ref{Details of bi-ImCsiNet}. After feature extraction and compression by the three-layer bi-LSTM network, an $L$-dimensional compressed codeword is obtained and sent to the quantization module to generate the corresponding PMI for feedback. Uniform quantization in (\ref{uniformquantization}) is adopted in the quantization module and the total feedback overhead is $L \times B$ bits. The decoder of bi-ImCsiNet is consistent with that of ImCsiNet-m, and the performance metric and loss function are the same as (\ref{rho_m}) and (\ref{loss_m}), respectively. Since LSTM shares parameters at each time step, the total parameters of bi-ImCsiNet are dramatically reduced compared with ImCsiNet-m composed of FC layers. Specifically, the parameter number of LSTM can be calculated by
\begin{equation}
\label{parameters_LSTM}
N_{\rm{LSTM}} = 4(O \times I + O^{2} + O),
\end{equation}
where $I$ and $O$ denote the input and the output dimensionality of each time step, respectively. Therefore, the parameter number of bi-LSTM can be expressed as $2N_{\rm{LSTM}}$. According to $L = \alpha \times 2N_{t}N_{s}$, we can get $M = \alpha \times 2N_{t}$. The parameters of the encoder in ImCsiNet-m are $(80+16\alpha)N_{s}^{2}N_{t}^{2}+(80+10\alpha)N_{s}N_{t}$ while the parameters of the encoder in bi-ImCsiNet are $(89+4\alpha^{2}+2\alpha)8N_{t}^{2}+(9+2\alpha)8N_{t}$. The parameter numbers of the two encoders are compared in Section \ref{results}.

\begin{table*}[tb]
\centering
\caption{Details of the encoder in bi-ImCsiNet}
\setlength{\tabcolsep}{5mm}{
\begin{tabular}{c|ccccc} \hline \hline
                         & Layer name & Output shape  & Activation function & Parameters \\
\hline \hline
\multirow{5}{*}{Encoder} & Input    & $N_{s} \times 2N_{t}$  & $\backslash$  & 0                            \\
                         & bi-LSTM1 & $N_{s} \times 8N_{t}$  & Tanh          & $8(80N_{t}^{2}+8N_{t})$      \\
                         & bi-LSTM2 & $N_{s} \times N_{t}$   & Tanh          & $8(9N_{t}^{2}+N_{t})$        \\
                         & bi-LSTM3 & $N_{s} \times M$       & Tanh          & $8(M \times N_{t}+M^{2}+M)$  \\ 
                         & Reshape  & $L \times 1$           & $\backslash$  & 0                            \\
\hline \hline
\end{tabular}}
\label{Details of bi-ImCsiNet}
\end{table*}

\section{Simulation Results and Discussions}
\label{results}
In this section, we conduct a numerical simulation to evaluate the performance of the proposed DL-based implicit feedback architectures. First, we elaborate on the simulation settings, including channel data generation and network training details. Then, we demonstrate the performance of ImCsiNet-s under different antenna configurations. Finally, we investigate the recovery performance and network complexity of the proposed ImCsiNet-m and bi-ImCsiNet for multiple RBs. The performance of Type I and Type II codebooks serves as a benchmark.

\subsection{Simulation Settings}
\subsubsection{Channel Data Generation}
The channel models adopted in our simulation are defined in 3GPP TR 38.901 \cite{3GPP2019channel}, which cover all the frequency bands and the bandwidth requirements of 5G. For the OFDM system with a single RB, the downlink channels are generated by the clustered delay line (CDL)-C channel model at 4\,GHz with a time delay of 300\,ns, which is a nonline-of-sight scenario with many scattering paths. The BS is configured with the cross-polarized single-panel antenna array with $N_{t}=8$ or $N_{t}=32$ transmit antennas, and the UE is deployed with $N_{r}=2$ receiver antennas. For $N_{t}=8$, the antenna port configuration $(N_1,N_2)$ is $(2,2)$. For $N_{t}=32$, the antenna port configuration $(N_1,N_2)$ is $(8,2)$. The oversampling factor $(O_1,O_2)$ is $(4,4)$ for both antenna configurations. Full channel matrix $\mathbf{H}$ is generated under the two settings, and then the corresponding eigenvectors $\mathbf{v}$ are calculated as input samples according to Section \ref{Narrowband}. The two datasets are denoted as CDL-C(8T2R) and CDL-C(32T2R), respectively. 

For the OFDM system with multiple RBs, the urban microcell (UMi) scenario is adopted to generate the downlink channels. The system has $N_{\rm{RB}}=52$ RBs in the frequency domain and the carrier frequency of the downlink is 3.5\,GHz. The entire band is divided into $N_{s}=13$ subbands, each of which contains four RBs. In such a scenario, the BS is configured with the cross-polarized single-panel antenna array with $N_{t}=32$ transmit antennas and the UE is deployed with $N_{r}=4$ receiver antennas. The antenna port configuration $(N_1,N_2)$ of BS is $(8,2)$ and the oversampling factor $(O_1,O_2)$ is $(4,4)$. We generate full channel matrices and then calculate the corresponding eigen matrices $\mathbf{V}_{\rm{stack}}$ as input samples according to Section \ref{Wideband}. The dataset is denoted as UMi(32T4R).

We generate 720,000 samples for each dataset. The samples are randomly shuffled and divided into training, validation, and test sets containing 80\%, 10\%, and 10\% samples, respectively.

\subsubsection{Training Details}
With supervised learning, the ideal output of the proposed NNs is the input sample, and thus the label of the input sample is set to the sample itself. The proposed DL-based architectures are built using Keras library \cite{Keras}, with Tensorflow \cite{Tensorflow} as the backend, and all simulation is implemented on an NVIDIA DGX-1 workstation. The adaptive moment estimation (Adam) optimizer \cite{Adam} is used to update the weights of NNs in the simulation. The batch size is set as 1,024 and the training epoch is 1,000. The initial learning rate is 0.001 and will be reduced by half if the validation loss does not decrease over 50 epochs. The loss function used in the training process and the performance metric for evaluating the trained model are clearly described in Section \ref{architecture}.

\subsection{NNs for Single RB}

\subsubsection{Performance of ImCsiNet-s}
We first use the CDL-C(8T2R) dataset to train ImCsiNet-s and evaluate the recovery performance. According to Type I codebook introduced in Section \ref{codebook}, the codeword, $\mathbf{w}_{\rm{I}}$, corresponds to the eigenvector, $\mathbf{v}$. When $N_{t}=8$ with $(N_1,N_2)=(2,2)$ and $(O_1,O_2)=(4,4)$, the total number of codewords is 256. Therefore, the feedback overhead required by Type I codebook is $8$ bits. The cosine similarity between Type I codebook-based codewords, $\mathbf{w}_{\rm{I}}$, and the original eigenvector, $\mathbf{v}$, is $0.8677$ and serves as the performance comparison benchmark. The recovery performance of the proposed ImCsiNet-s, i.e., the cosine similarity $\rho_{\rm{s}}$ between the recovered $\hat{\mathbf{v}}$ and the original $\mathbf{v}$, with different feedback bits $N_{\rm{bits}}$ is shown in Table \ref{ImCsiNet-s(8T2R)}. 

\begin{table*}[tb]
\centering
\caption{The recovery performance of ImCsiNet-s in CDL-C(8T2R) dataset}
\setlength{\tabcolsep}{4.5mm}{
\begin{tabular}{c|ccccccc|c} \hline \hline
\multicolumn{1}{c|}{Schemes}  & \multicolumn{7}{c}{ImCsiNet-s}  & \multicolumn{1}{|c}{Type I}  \\
\hline
$N_{\rm{bits}}$ (bits) & 2      & 3      & 4      & 5      & 6      & 7      & 8      & 8  \\
$\rho_{\rm{s}}$        & 0.7529 & 0.7856 & 0.8199 & 0.8464 & $\textbf{0.8685}$ & 0.8808 & 0.8922 & 0.8677 \\
\hline \hline
\end{tabular}}
\label{ImCsiNet-s(8T2R)}
\end{table*}

\begin{figure*}[tb]
    \centering
     \subfigure[PSE for $N_{t}=8$]{
     \label{PSE:a}
     \includegraphics[width=0.444\textwidth,trim=5 6 12 4,clip]{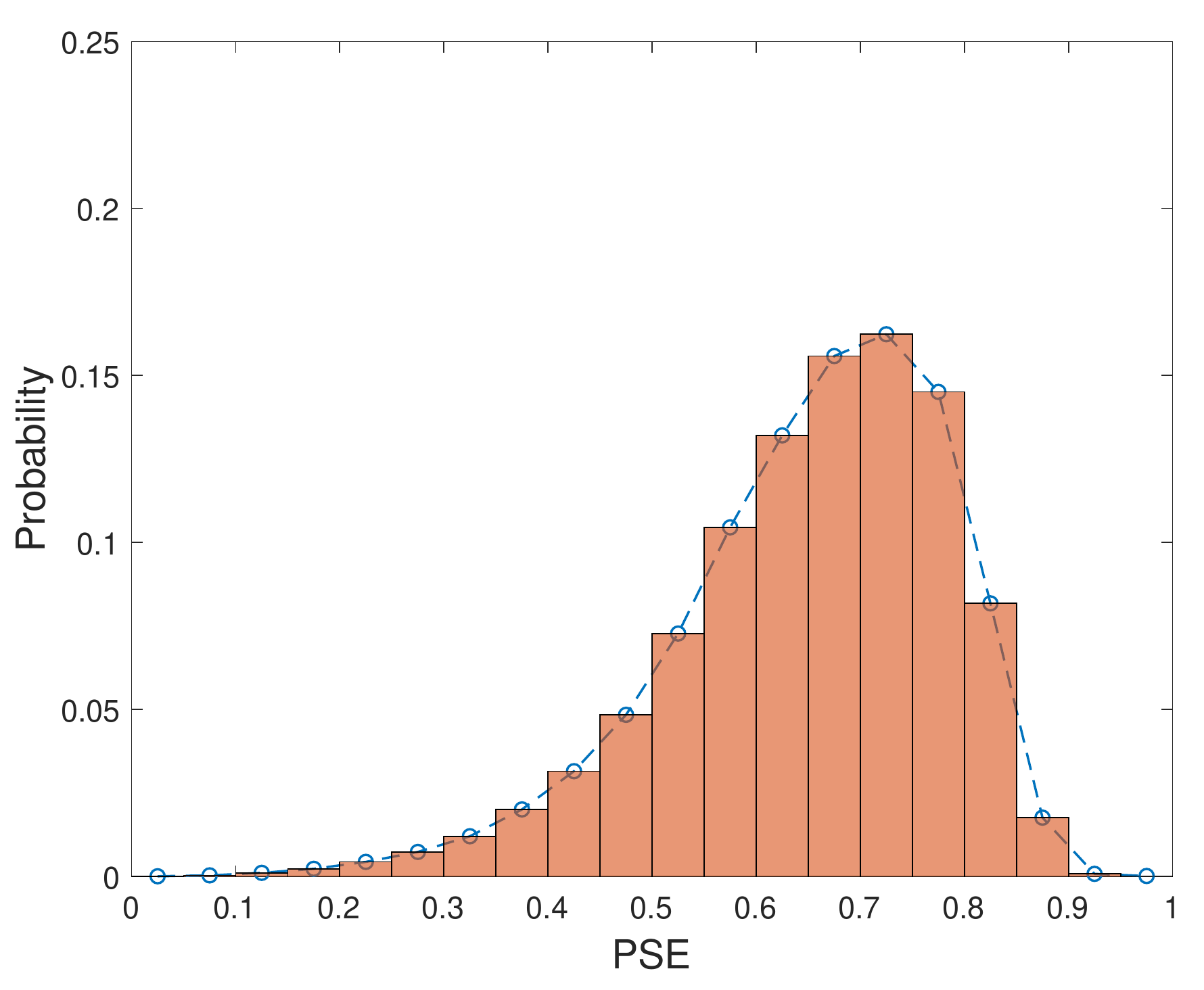} 
     }
    \quad
     \subfigure[PSE for $N_{t}=32$]{
     \label{PSE:b}
     \includegraphics[width=0.45\textwidth,trim=10 8 12 4,clip]{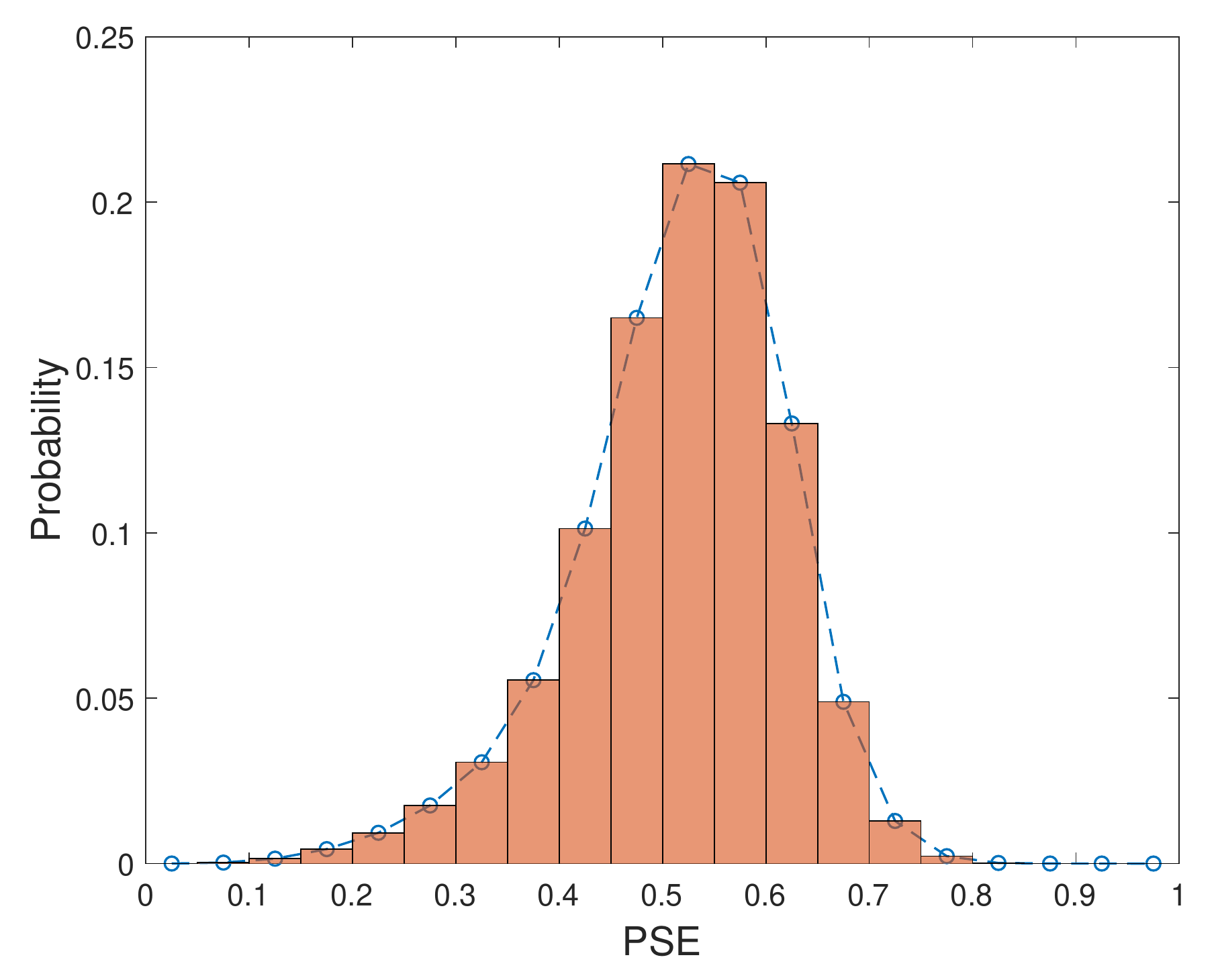} 
     }
	\caption{PSE distribution of eigenvectors for $N_{t}=8$ and $N_{t}=32$. PSE is a metric for the compressibility of eigenvectors.} 
	\label{quantization bit}
\end{figure*}

From the table, when the feedback overhead of ImCsiNet-s is $6$ bits, the cosine similarity is $0.8685$, which has been superior to the performance of the Type I codebook-based feedback scheme. The corresponding feedback bits of ImCsiNet-s are reduced to $75.0\%$ of that of Type I codebook. The performance advantage of the DL-based implicit feedback scheme is not particularly significant. The possible reason is that the freedom degree of the channel is small and the compressibility is not strong. The PSE distribution of the eigenvectors for $N_{t}=8$ is shown in Fig. \ref{PSE:a}. From the figure, the PSE is mainly distributed between $0.55$ and $0.85$. The distribution is diffused and the average PSE of all samples is $0.6511$, which increases the difficulty of learning an optimal compression strategy for all samples. 

By contrast, the existing DL-based explicit feedback architecture, e.g., CsiNet, transforms the full channel matrix into a sparse domain and then extracts features to compress it into a low-dimensional codeword. However, the eigenvectors to be fed back in the implicit scheme are extracted from full channel matrices, which is equivalent to a process of feature extraction and dimension compression. Therefore, the eigenvectors have fewer operable degrees of freedom and are not conducive to compression. In the subsequent simulation, feedback performance is measured by increasing the number of transmit antennas and the number of RBs.

\subsubsection{Effect of Transmit Antenna Number}
The channel vector is approximately sparse in the angular domain and the sparsity increases with the number of the transmit antennas at the BS \cite{Wen2015Channel}. Inspired by the property, we study the effect of transmit antenna numbers on the DL-based implicit feedback scheme. In the experiment, we increase the transmit antenna number from $N_{t}=8$ to $N_{t}=32$, and generate the CDL-C(32T2R) dataset to train and test ImCsiNet-s. Similarly, the Type I codebook-based $\mathbf{w}_{\rm{I}}$ corresponding to the eigenvectors are calculated as a performance benchmark of ImCsiNet-s. With $(N_1,N_2)=(8,2)$ and $(O_1,O_2)=(4,4)$, the total number of $\mathbf{w}_{\rm{I}}$ is 1024; thus the feedback overhead required by Type I codebook is $10$ bits. The cosine similarity between the obtained $\mathbf{w}_{\rm{I}}$ and the original eigenvector $\mathbf{v}$ is $0.7589$. As the number of the transmit antennas increases, the required number of feedback bits of the codebook increases while the recovery performance decreases, demonstrating that the increase of antennas introduces difficulty to the design of a codebook-based feedback scheme. As a comparison, the great recovery performance of ImCsiNet-s with different feedback bits $N_{\rm{bits}}$ is shown in Table \ref{ImCsiNet-s(32T2R)}.

\begin{table*}[tb]
\centering
\caption{The recovery performance of ImCsiNet-s in CDL-C(32T2R) dataset}
\setlength{\tabcolsep}{2.9mm}{
\begin{tabular}{c|ccccccccc|c} \hline \hline
\multicolumn{1}{c|}{Schemes}  & \multicolumn{9}{c}{ImCsiNet-s}  & \multicolumn{1}{|c}{Type I}  \\
\hline
$N_{\rm{bits}}$ (bits) & 2      & 3      & 4      & 5      & 6      & 7      & 8      & 9      & 10      & 10  \\
$\rho_{\rm{s}}$        & 0.4948 & 0.5910 & 0.6876 & 0.7350 & $\textbf{0.7726}$ & 0.7972 & 0.8322 & 0.8443 & 0.8557 & 0.7589 \\
\hline \hline
\end{tabular}}
\label{ImCsiNet-s(32T2R)}
\end{table*}

From Table \ref{ImCsiNet-s(32T2R)}, when the feedback overhead of ImCsiNet-s is reduced to $60.0\%$ of that of Type I codebook, ImCsiNet-s still outperforms the Type I codebook-based scheme. The gap between the feedback overhead of ImCsiNet-s and that of Type I codebook increases with the number of the transmit antennas. We also compare the compression performance with similar cosine similarity in the two antenna configurations. For example, when $N_{t}=8$ and $N_{\rm{bits}}=5$, the cosine similarity $\rho_{\rm{s}}$ is $0.8464$; when $N_{t}=32$ and $N_{\rm{bits}}=9$, the cosine similarity $\rho_{\rm{s}}$ is $0.8443$. The compression ratio $\alpha$ of the former is $5/16$ while that of the latter is $9/64$, which is significantly reduced. Under the similar requirement of cosine similarity, the eigenvectors with $N_{t}=32$ transmit antennas are more compressible, which is consistent with the conclusion obtained in \cite{Wen2015Channel}. As shown in Fig. \ref{PSE:b}, the PSE distribution of the eigenvectors for $N_{t}=32$ is mainly concentrated in the range of $0.40$ to $0.65$. The average PSE of all samples is $0.5168$. The PSE distribution for $N_{t}=32$ is relatively concentrated compared with that for $N_{t}=8$ and the average PSE is smaller, indicating stronger compressibility. The analysis of the PSE value is consistent with the simulation results. Therefore, the increase of antenna dimension can improve compressibility of the eigenvectors to some extent, which makes the advantage of DL-based implicit feedback more obvious than the codebook-based scheme.

\subsection{NNs for Multiple RBs}
\label{Wideband analysis}

\subsubsection{Effect of Quantization Bits}
The quantization module of ImCsiNet-s adopts binarization operation, which is equivalent to a 1-bit quantization. Thus, a trade-off between the number of quantization bits and the compressed dimension in ImCsiNet-s is not necessary. However, for multiple RBs, $B$-bit uniform quantization is adopted in the proposed ImCsiNet-m and bi-ImCsiNet. With the same feedback overhead, $N_{\rm{bits}}$, different combinations of the number of quantization bits, $B$, and the total compressed dimension, $L$, can be used. By substituting $L = \alpha \times 2N_{t}N_{s}$ into (\ref{feedback_bits}), the correspondence between the number of quantization bits, $B$, and the compression ratio, $\alpha$, can be obtained. When the feedback overhead is fixed, $\alpha$ decreases with the increase of $B$ while $\alpha$ increases with the decrease of $B$. For example, when the total feedback overhead is $N_{\rm{bits}}=312$ bits, the corresponding $L$ is $156$ and the compression ratio $\alpha$ is $3/16$ if $B=2$; the corresponding $L$ is $52$ and the compression ratio $\alpha$ is $1/16$ if $B=6$. Therefore, carefully selecting the number of quantization bits to achieve the optimal recovery performance with the same $N_{\rm{bits}}$ is necessary. The dataset used in the simulation is UMi(32T4R). We fix the number of feedback bits, $N_{\rm{bits}}$, as 156, 208, 312, and 624, and measure the reconstruction performance of ImCsiNet-m and bi-ImCsiNet, respectively, with different values of $B$. The cosine similarity between the recovered eigen matrix $\hat{\mathbf{V}}_{\rm{stack}}$ and the original eigen matrix $\mathbf{V}_{\rm{stack}}$ is shown in Fig. \ref{quantization bit}. When $B \geq 2$, the recovery performance becomes worse with increasing $B$ under a fixed number of feedback bits. Compared with reducing the number of quantization bits, the performance loss caused by decreasing the compression ratio is larger, indicating that the recovery performance of the proposed NNs is more sensitive to the compression ratio change. Moreover, whether for ImCsiNet-m or bi-ImCsiNet, the recovery performance is the best when the number of quantization bits is 2 with a fixed number of feedback bits. Therefore, $B=2$ is set for subsequent simulation. 

\begin{figure*}[tb]
    \centering
     \subfigure[Effect of quantization bits on ImCsiNet-m]{
     \label{quantization bit:a}
     \includegraphics[width=0.46\textwidth,trim=8 6 10 2,clip]{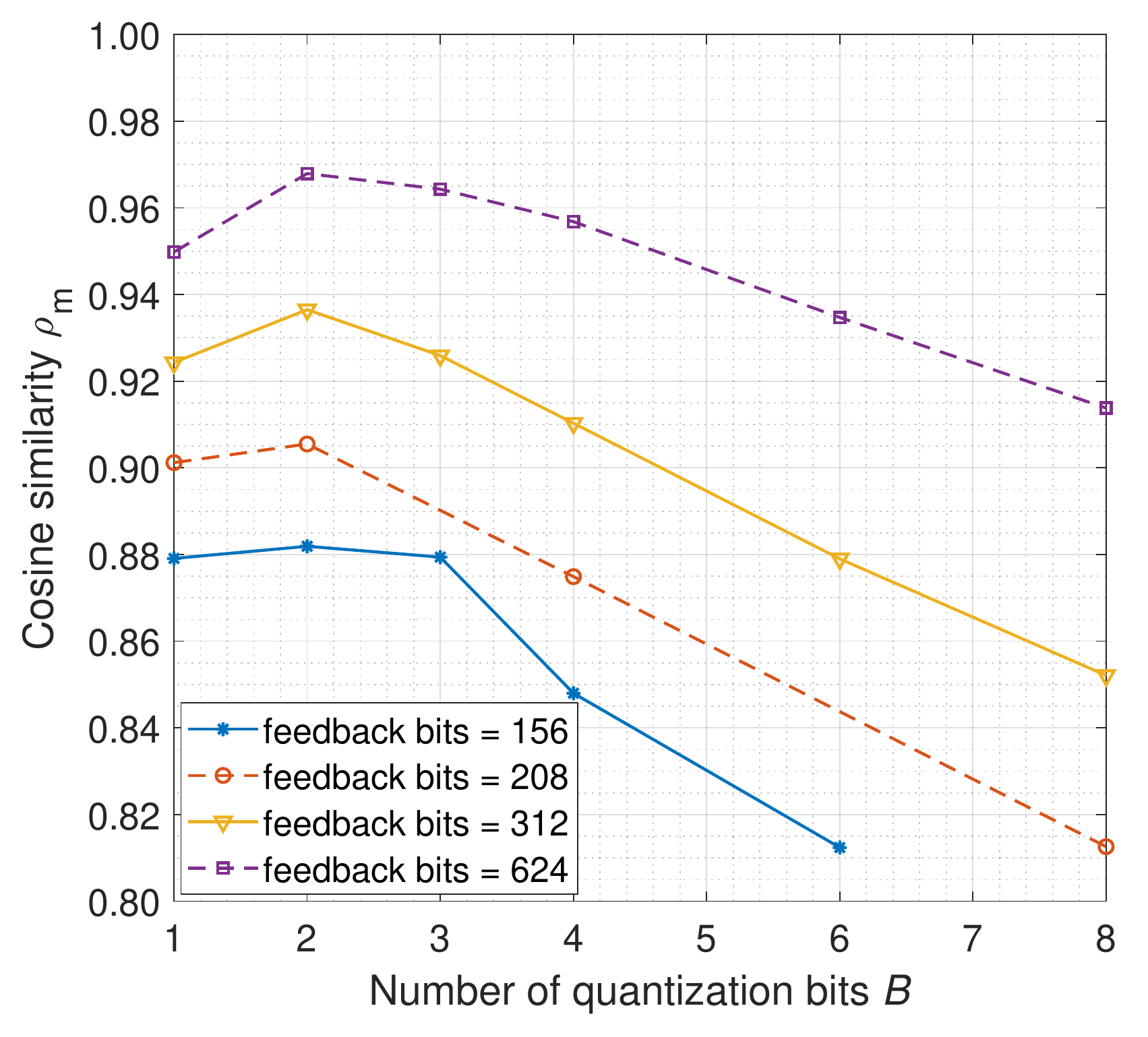}
     }
    \quad
     \subfigure[Effect of quantization bits on bi-ImCsiNet]{
     \label{quantization bit:b}
     \includegraphics[width=0.453\textwidth,trim=8 6 9.5 2,clip]{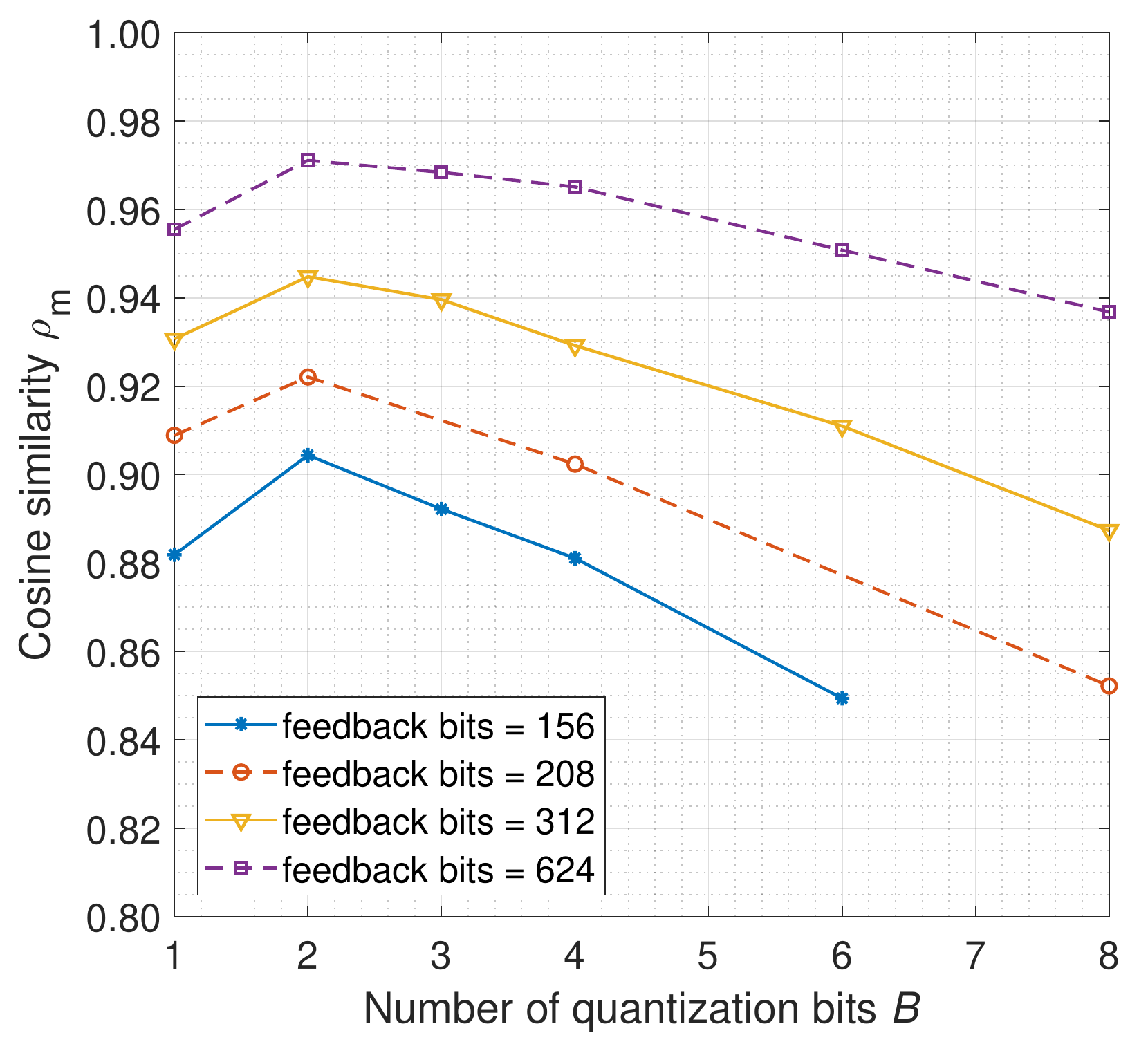} 
     }
	\caption{Recovery performance using different numbers of quantization bits, $B$, under a fixed number of feedback bits. The UMi(32T4R) dataset is used for simulation. Four fixed feedback bits are set, which are 156, 208, 312 and 624, respectively.} 
	\label{quantization bit}
\end{figure*}

\subsubsection{Performance Comparison between ImCsiNet-m and bi-ImCsiNet}
For multiple RBs, we have proposed two network structures in Section \ref{architecture for wideband}, i.e., ImCsiNet-m and bi-ImCsiNet. In this experiment, we evaluate the performance of the two NNs from several aspects, including cosine similarity, parameter complexity, and time complexity. The UMi(32T4R) dataset is used for simulation. In the wideband system that implements subband feedback, Type II codebook combines the wideband and subband information to generate a codebook, which has a higher resolution than Type I codebook. Therefore, to further improve the gain of DL-based implicit feedback, the recovery performance of Type II codebook with high precision is used as the comparison benchmark for the proposed NNs. The Type II codebook-based $\mathbf{W}_{\rm{II}}$ corresponding to eigen matrix $\mathbf{V}_{\rm{stack}}$ in the UMi(32T4R) dataset is generated according to Section \ref{codebook}. The cosine similarity between $\mathbf{W}_{\rm{II}}$ and the original $\mathbf{V}_{\rm{stack}}$ is $0.9042$ and the approximate feedback overhead is 300 bits. The recovery performance of ImCsiNet-m and bi-ImCsiNet, i.e., the cosine similarity $\rho_{\rm{m}}$ between the recovered $\hat{\mathbf{V}}_{\rm{stack}}$ and the original $\mathbf{V}_{\rm{stack}}$, with different feedback bits $N_{\rm{bits}}$ is shown in Table \ref{ImCsiNet-m&bi-ImCsiNet(32T4R)}. 

\begin{table*}[tb]
\centering
\caption{Recovery performance of ImCsiNet-m and bi-ImCsiNet in UMi(32T4R) dataset}
\begin{threeparttable}
\setlength{\tabcolsep}{5mm}{
\begin{tabular}{c|ccc} \hline \hline
Schemes                      & $N_{\rm{bits}}$ (bits) & $N_{\rm{ps}}$\tnote{*} (bits) & $\rho_{\rm{m}}$  \\
\hline\hline
Type II                      & 300    & $\approx23$  & 0.9042  \\
\hline
\multirow{8}{*}{ImCsiNet-m}  & 832    & 64    & 0.9747  \\
                             & 624    & 48    & 0.9679  \\
                             & 416    & 32    & 0.9523  \\
                             & 312    & 24    & 0.9365  \\
                             & 208    & 16    & \textbf{0.9055}  \\
                             & 156    & 12    & 0.8819  \\
                             & 104    & 8     & 0.8428  \\
                             & 78     & 6     & 0.8153  \\
\hline
\multirow{8}{*}{bi-ImCsiNet} & 832    & 64    & 0.9769  \\
                             & 624    & 48    & 0.9711  \\
                             & 416    & 32    & 0.9574  \\
                             & 312    & 24    & 0.9448  \\
                             & 208    & 16    & 0.9221  \\
                             & 156    & 12    & \textbf{0.9044}  \\
                             & 104    & 8     & 0.8721  \\
                             & 78     & 6     & 0.8507  \\
\hline \hline
\end{tabular}}
\begin{tablenotes}
\footnotesize
\item[*] $N_{\rm{ps}}$ represents the number of feedback bits per subband, as defined in \ref{architecture for wideband}. The total feedback overhead satisfies $N_{\rm{bits}}=N_{s} \times N_{\rm{ps}}$.
\end{tablenotes}
\end{threeparttable}
\label{ImCsiNet-m&bi-ImCsiNet(32T4R)}
\end{table*}

From the table, when total feedback overhead $N_{\rm{bits}}$ of ImCsiNet-m is $208$ bits, the corresponding cosine similarity $\rho_{\rm{m}}$ is equivalent to that of Type II codebook. The number of feedback bits is $208$, which is approximately $69.3\%$ of the feedback overhead using Type II codebook. When $N_{\rm{bits}}$ of bi-ImCsiNet is $156$ bits, corresponding $\rho_{\rm{m}}$ is equivalent to that of Type II codebook. The feedback overhead is $52.0\%$ of that using Type II codebook. Therefore, compared with the Type II codebook-based feedback scheme, the performance of the two proposed NNs is significantly improved. Moreover, bi-ImCsiNet outperforms ImCsiNet-m in terms of $\rho_{\rm{m}}$ under all feedback overhead, indicating that the introduction of bi-LSTMs in the encoder to extract the subband correlation is of considerable importance to improving the recovery performance of the NNs and the improvement effect is more obvious under low feedback bits.

\begin{figure*}[tb]
    \centering 
     \includegraphics[width=0.5\textwidth,trim=7 6 9 6,clip]{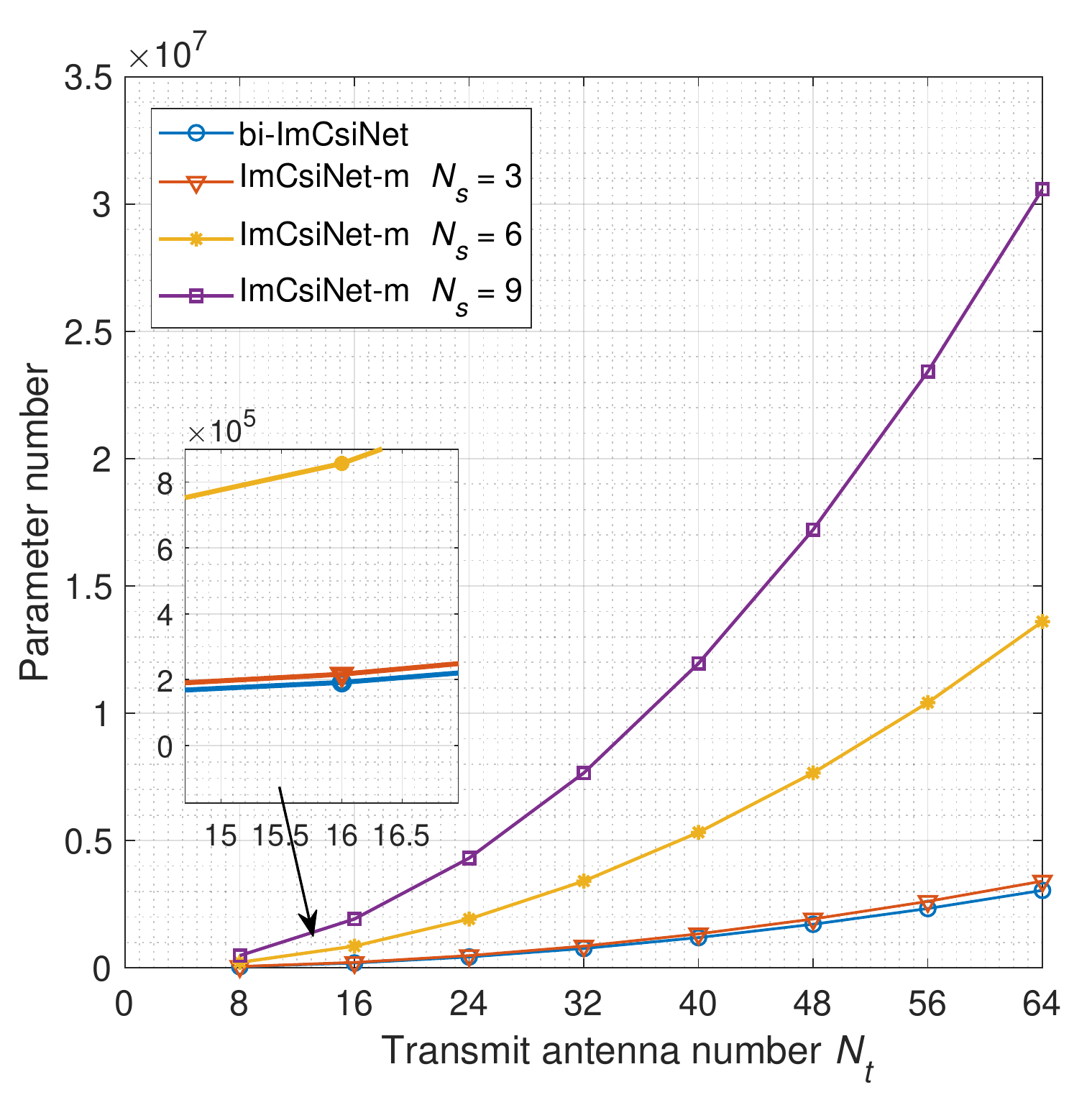}    
	\caption{\label{parameter complexity} Parameter numbers of the encoders in ImCsiNet-m and bi-ImCsiNet. The subband number, $N_{s}$, only affects the parameter number of the encoder in ImCsiNet-m. } 
\end{figure*}

We also compare the parameter complexity of ImCsiNet-m and bi-ImCsiNet. Given that the decoders in the two NNs are the same, the only difference is that the encoder in ImCsiNet-m is composed of three FC layers while the encoder in bi-ImCsiNet is composed of a three-layer bi-LSTM network. Therefore, we only need to compare the parameter numbers of the two encoders. As analyzed in Section \ref{architecture for wideband}, the parameters of the encoders in ImCsiNet-m and bi-ImCsiNet can be expressed as $(80+16\alpha)N_{s}^{2}N_{t}^{2}+(80+10\alpha)N_{s}N_{t}$ and $(89+4\alpha^{2}+2\alpha)8N_{t}^{2}+(9+2\alpha)8N_{t}$, respectively. They depend on three variables, i.e., compression ratio $\alpha$, subband number $N_{s}$, and transmit antenna number $N_{t}$. The compression ratio $\alpha$ is a value in the range of $(0,1)$. The number of subbands, $N_{s}$, only affects the parameter number of the encoder in ImCsiNet-m. In 3GPP TS 38.214 when the physical RB occupied by the bandwidth reaches 24, the whole bandwidth can be divided to form subbands and the minimum subband number is 3 \cite{3GPP2019codebook}. Therefore, given the minimum value of $N_{s}$, the parameter number of ImCsiNet-m can be scaled, i.e., $(80+16\alpha)N_{s}^{2}N_{t}^{2}+(80+10\alpha)N_{s}N_{t} \geq (80+16\alpha)9N_{t}^{2}+(80+10\alpha)3N_{t}$. Then, we subtract the parameter number of bi-ImCsiNet from the scaled parameter number of ImCsiNet-m and get $(-32\alpha^{2}+128\alpha+8)N_{t}^{2}+(14\alpha+168)N_{t}$. Given that $\alpha \in (0,1)$, both coefficients $(-32\alpha^{2}+128\alpha+8)$ and $(14\alpha+168)$ are positive. In other words, we can obtain $(80+16\alpha)9N_{t}^{2}+(80+10\alpha)3N_{t} \geq (89+4\alpha^{2}+2\alpha)8N_{t}^{2}+(9+2\alpha)8N_{t}$. As a result, the parameter number of ImCsiNet-m is larger than that of bi-ImCsiNet. 

The value of compression ratio $\alpha$ has no effect on the comparison. To directly observe the influence of subband number and transmit antenna number on the parameters, the compression ratio can be set as a specific value. Subband number $N_{s}$ is set as 3, 6, and 9, respectively. The curves of the encoder parameters changing with transmit antenna number $N_{t}$ are shown in Fig. \ref{parameter complexity}. From the figure, the parameters of the encoder in ImCsiNet-m significantly increase with the increase of $N_{s}$. When $N_{s}=3$, the parameter number of ImCsiNet-m is still more than that of bi-ImCsiNet, illustrating that bi-ImCsiNet has lower parameter complexity than ImCsiNet-m. The curve trend is consistent with the theoretical analysis. In addition, as the transmit antenna number, $N_{t}$, grows, the parameters of both ImCsiNet-m and bi-ImCsiNet increase dramatically, resulting in huge storage overhead at the UE. Therefore, further optimization design of NNs can be considered in the future. Finally, we compare the time complexity of the two NNs. The runtime of ImCsiNet-m is approximately $0.0001$\,s while the runtime of bi-ImCsiNet is approximately $0.001$\,s. In conclusion, bi-ImCsiNet obviously outperforms ImCsiNet-m in terms of recovery performance and parameter complexity at the cost of a slight increase in runtime.

\section{Conclusion}
\label{Conclusion}
In this paper, we proposed a novel DL-based implicit feedback architecture, using NNs to replace the PMI encoding module at the UE and the PMI decoding module at the BS, respectively. The core idea was to retain the framework of implicit feedback mechanism in the existing standards and use DL to enhance implicit feedback.

First, we designed ImCsiNet-s composed of FC layers for the OFDM system with a single RB. The eigenvectors extracted from full channel matrices were compressed and quantized by the encoder to generate the corresponding PMI for feedback and were recovered by the decoder for downlink transmission. Then, we extended the implicit framework to multiple RBs and built ImCsiNet-m to compress and recover the eigen matrix stacked by the eigenvectors of each subband. On this basis, we used a bi-LSTM network to replace the FC layers of the encoder in ImCsiNet-m. The new structure, bi-ImCsiNet, was designed to extract the correlation of the eigenvectors among different subbands. Simulation results have demonstrated that the proposed DL-based implicit feedback architecture outperforms Type I and Type II codebooks in 5G NR, potentially reducing the feedback overhead while ensuring recovery performance. In addition, bi-ImCsiNet is better than ImCsiNet-m in terms of recovery similarity and parameter complexity, indicating that the extraction of subband correlation can help improve the feedback performance. The proposed DL-based implicit feedback scheme has provided a new direction for future CSI feedback research.

\bibliographystyle{IEEEtran}
\bibliography{IEEEabrv,reference}
\end{document}